\documentclass[a4paper,12pt]{article}
\pdfoutput=1 

\usepackage{amsmath,amssymb}
\usepackage[pdftex]{graphicx}
\usepackage[pdftex]{hyperref}
\hypersetup{colorlinks=true,linkcolor=blue,urlcolor=blue,filecolor=blue,citecolor=blue}
\usepackage{epsf}
\usepackage{cite}
\usepackage{physics}
\usepackage{fancyhdr}
\usepackage{enumerate} 
\usepackage[usenames]{xcolor}
\usepackage{tikz}
\usetikzlibrary{snakes}
 \usetikzlibrary{shapes.misc}

\makeatletter

\newcommand{\xv}{{\mathbf x}}
\newcommand{\zh}{{\hat{z}}}

\g@addto@macro\bfseries{\boldmath}

\def\refchecklabelfontsize{\fontsize{5pt}{5pt}\selectfont}
\let\mark@size=\refchecklabelfontsize

\def\half{{\frac{1}{2}}}

\def\p{\partial}

\def\unit{{1\kern-.65ex {\rm l}}}
\def\1{{1\kern-.65ex {\rm l}}}

\def\Im{\mathop{\mathrm{Im}}\nolimits}

\def\Re{\mathop{\mathrm{Re}}\nolimits}

\def\sign{\mathop{\mathrm{sign}}\nolimits}

\let\ev=\bracket

\def\Omegah{{\widehat{\Omega}}}

\def\htilde{{\widetilde{h}}}

\def\zb{{\overline{z}}}

\def\Fb{{\overline{F}}}
\def\Gb{{\overline{G}}}

\def\cC{{\cal C}}

\def\cG{{\cal G}}
\def\cH{{\cal H}}

\def\cN{{\cal N}}
\def\cO{{\cal O}}
\def\cP{{\cal P}}
\def\cQ{{\cal Q}}

\def\bbC{{\mathbb{C}}}

\def\bbR{{\mathbb{R}}}

\def\bbZ{{\mathbb{Z}}}

\newcount\hour \newcount\minute
\hour=\time \divide \hour by 60
\minute=\time
\def\now{%
\ifnum \hour<13
  \ifnum \hour=0 \advance \hour by 12 \number\hour:\else \number\hour:\fi%
     \ifnum \minute<10 0\fi%
     \number\minute%
\ A.M.%
\else \advance \hour by -12 \number\hour:%
  \ifnum \minute<10 0\fi%
  \number\minute%
  \ P.M.%
\fi%
}

\makeatother

\begin{document}

\baselineskip=18pt  
\numberwithin{equation}{section}  

\renewcommand{\headrulewidth}{0pt}
%



\thispagestyle{empty}

\vspace*{-2cm} 
\begin{flushright}
YITP-23-171\\
\end{flushright}

\vspace*{2.5cm} 
\begin{center}
 {\LARGE Exact Non-Abelian Supertubes}\\
 \vspace*{1.7cm}
 Ryo Nemoto$^1$ and Masaki Shigemori$^{1,2}$\\
 \vspace*{1.0cm} 
$^1$\,Department of Physics, Nagoya University\\
Furo-cho, Chikusa-ku, Nagoya 464-8602, Japan\\[1ex]
$^2$\,Center for Gravitational Physics,\\
Yukawa Institute for Theoretical Physics, Kyoto University\\
Kitashirakawa Oiwakecho, Sakyo-ku, Kyoto 606-8502, Japan
\end{center}
\vspace*{1.5cm}

\noindent
Supertubes are supersymmetric configurations in string theory in which
branes are extending along a closed curve.  For a supertube of
codimension two, its dipole charge is characterized by the duality
monodromy around the closed curve.  When multiple codimension-2
supertubes are present, the monodromies around different supertubes can
be non-commuting, namely non-Abelian.  Non-Abelian configurations of
supertubes are expected to play an important role in non-perturbative
physics of string theory, especially black holes.  In this paper, in the
framework of five-dimensional supergravity, we construct exact solutions
describing codimension-2 supertubes in three-dimensional space.  We use
an extension formula to construct a three-dimensional solution from a
two-dimensional seed solution.  The two-dimensional seed is an F-theory
like configuration in which a torus is nontrivially fibered over a
complex plane.  In the first example, there is a stack of circular
supertubes around which there is a non-trivial monodromy.  In some cases
this can be thought of as a microstate of a black hole in AdS$_2\times
S^2$.  The second example is an axi-symmetric solution with two stacks
of circular supertubes with non-Abelian monodromies.  In addition, there
is a continuous distribution of charges on the symmetry axis.

\newpage
\setcounter{page}{1} 



\tableofcontents


\newpage
\section{Introduction}

Black-hole microstates pose a long-standing puzzle in theoretical
physics.  Having thermodynamical entropy, black holes must represent
ensembles of many microstates, but the gravity picture of those
individual microstates remains poorly understood.  Some of the
black-hole microstates are known to be realized as smooth horizonless
solutions of classical supergravity, called microstate
geometries. Because
supergravity is the low-energy effective theory of string theory,
microstate geometries provide a unique, top-down approach to studying
the microscopic nature of individual microstates.

The known microstate geometries roughly fall into two categories.  The
first one is the multi-center solution \cite{Behrndt:1997ny,
Gauntlett:2002nw, Bates:2003vx, Bena:2004de, Gauntlett:2004qy,
Bena:2005va, Berglund:2005vb, Meessen:2006tu}, which is a solution of
four- or five-dimensional supergravity and represents bound states of
branes wrapping internal cycles.  The second one is the superstratum
\cite{Bena:2015bea, Shigemori:2020yuo}, which is a solution of
six-dimensional supergravity and contains traveling waves along a sixth
direction.  There are both supersymmetric and non-supersymmetric
versions to these solutions, but we restrict ourselves to the
supersymmetric solutions which are in better control.  The solutions in
these two categories both have a large entropy, but it is parametrically
smaller than the entropy of the black hole, of which these solutions are
microstates \cite{deBoer:2009un, Shigemori:2019orj,
Mayerson:2020acj}.  Therefore, these solutions correspond to atypical
microstates in the relevant black-hole ensembles.

To go beyond these existing microstate geometries and find (more)
typical microstates, there are multiple possible directions.  One
direction is to include full string-theory physics. Typical microstates
must certainly be describable in string theory, although it is not
completely obvious if full-fledged string theory is necessary for
supersymmetric microstates.  Another direction is to remain in
supergravity but go to higher dimensions by including dependence on more
internal directions; see \cite{Bena:2022wpl, Bena:2022fzf, Bena:2023rzm,
Eckardt:2023nmn, Bena:2023fjx} for recent work in that direction.  Still
another direction is to remain in the same supergravity setup of the
known microstate geometries but look for more general solutions.
In fact, in multi-center solutions, one normally considers brane sources
(or centers) that have codimension three, but there are more general
solutions with codimension-2 brane sources \cite{Park:2015gka,
Fernandez-Melgarejo:2017dme, Shigemori:2021pir}. This extended class of
solutions must lead to a larger entropy.

In this paper, we continue the study of multi-center solutions with
codimension-2 centers initiated in \cite{Park:2015gka,
Fernandez-Melgarejo:2017dme}.  Multi-center solutions are supersymmetric
solutions of four- or five-dimensional supergravity obtained by
compactifying type IIA string theory or M-theory on a Calabi-Yau 3-fold
$X$. The solution is completely characterized by a set of real harmonic
functions on a flat spatial $\bbR^3$ base \cite{Gauntlett:2004qy}:
\begin{align}
 (V,K^I,L_I,M)\equiv \cH,\qquad \bigtriangleup \cH=0,\label{lozb5Dec23}
\end{align}
where $I=1,\dots,n$ with $n=h^{1,1}(X)$, and
$\bigtriangleup=\sum_{i=1}^3(\p/\p x_i)^2$.  As harmonic functions, $\cH$
can have codimension-3 (pointlike) sources.  Codimension-3 sources in
$V$, $K^I$, $L_I$, and $M$ correspond to D6, D4, D2 and D0-branes, respectively,
sitting at the location of the source and wrapping cycles in $X$.  In
this paper, we focus on $X=T^6$ regarded as $T^2_{45}\times
T^2_{67}\times T^2_{89}$ (the so-called STU model), and therefore
$I=1,2,3$ ($n=3$).  The harmonic function $\cH$ can also have
codimension-2 source along some curve $\cC$ inside $\bbR^3$.  A new
feature with such codimension-2 sources is that, as we go around $\cC$,
the harmonic functions can undergo a monodromy transformation,
\begin{align}
 \cH\to M \cH,\label{gnzz5Dec23}
\end{align}
where $M$ is an $(2n+2)\times (2n+2)$ matrix representing a $U$-duality
transformation (for $X=(T^2)^3$, the $U$-duality group is
$SL(2,\bbZ)^3$).  In string theory, codimension-2 sources correspond to
branes whose worldvolume extends along $\cC$ and, as we go around such a
brane, we generally undergo a duality transformation.  For example, if
we have $k$ NS5-branes along $\cC$ (with other four spatial directions
wrapping internal directions), a component of the NS-NS $B$-field jumps
by $k$ units as we go around them. This is a part of $T$-duality
transformation.  Because $\cH$ contains the $B$-field, we have a
monodromy that can be written as \eqref{gnzz5Dec23}.  Such codimension-2 sources are not mathematical curiosity but very
naturally arise in string they due to the supertube transition
\cite{Mateos:2001qs, Emparan:2001ux}.  For example, if we have D2(45)
and D2(67) branes put together, where 45 and 67 mean wrapped spatial
directions, they undergo supertube transition and generically polarize
into NS5-branes along 4567 and some other closed curve
$\cC\subset\bbR^3$.\footnote{We can also have D4(4567)-brane dipole
charge distributed along $\cC$.  There can be other dipole charges but
we do not consider them because they will break the symmetry of
$T^6=T^2_{45}\times T^2_{67} \times T^2_{89}$ that we assume.}  So, we
can generally regard codimension-2 sources as \emph{supertubes},
representing the bound states of codimension-3 branes.  We will discuss
this NS5-brane example in more detail later.  As another example, the
bound state of D4(6789) and D4(4589) branes is the so-called
$5^2_2$-brane \cite{Obers:1998fb, deBoer:2012ma}, extending along
$\cC\in\bbR^3$, and provides a different type of codimension-2 source.
Thus, when we study bound states in situations with various branes,
taking into account supertube transitions and codimension-2 sources with
non-trivial monodromies around them is natural and actually seems
unavoidable.

When multiple codimension-2 sources are present, say along $\cC_1$ and
$\cC_2$, it is possible that their monodromy matrices do not commute,
$[M_1,M_2]\neq 0$.  When this happens, we say that the configuration is
\emph{non-Abelian}.  For example, if NS5 and $5^2_2$-branes coexist,
their monodromy matrices do not commute and thus the configuration is
non-Abelian.  In this paper, we will especially be interested in
constructing multi-center solutions of non-Abelian supertubes.  In
\cite{Fernandez-Melgarejo:2017dme}, some multi-center solutions
involving non-Abelian supertubes were constructed in a perturbative
approach.  However, because of the perturbative nature, certain physical
aspects remained unclear, such as the behavior of fields away from the
supertubes.  In this paper, we will construct examples of exact
solutions of non-Abelian supertubes based on the techniques developed
there and clarify their physical aspects, such as charge distributions
and absence of closed timelike curves.

When constructing explicit solutions with codimension-2 branes, rather
than being general, we focus on the special class of solutions in which
the harmonic functions are written in terms of complex harmonic
functions $F,G$ as
\begin{align}
\begin{gathered}
  K^1=K^2=-\Im G,\qquad L_1=L_2=\Im F,\\
 L_3=V=\Re G,\qquad K^3=-2M=\Re F.
\end{gathered}\label{mnrg5Dec23}
\end{align}
Solutions of this restricted class are called SWIP solutions
\cite{Bergshoeff:1996gg}.  For this class, the duality group reduces to
$SL(2,\bbZ)$ and the harmonic functions $(F,G)$ transform as a doublet
as we go around codimension-2 branes; namely,
\begin{align}
\mqty(F\\ G)
\to M\mqty(F\\ G),\qquad
 M=\mqty(a&b\\c&d)\in SL(2,\bbZ),
\end{align}
where $a,b,c,d\in\bbZ$ and $ad-bc=1$.  The concrete $SL(2,\bbZ)$ matrix
depends on the specific codimension-2 brane that we consider.  So, the
problem of constructing solutions with codimension-2 branes amounts to
the problem of finding a pair of 3D harmonic functions $(F,G)$ with the
prescribed monodromies.  We construct such harmonic functions by
starting with a pair of 2D harmonic functions $(f,g)$ with non-trivial
monodromies, and extending them to 3D harmonic functions.

Our first example contains a single stack of $N$ codimension-2, circular
branes around which there is a nontrivial $SL(2,\bbZ)$ monodromy.
Although $(F,G)$ are nontrivial functions of the coordinates of
$\bbR^3$, the ratio $\tau=F/G$ is constant.  If $N\le 6$, this solution
is free from closed timelike curves (CTCs) and represent a new
microstate of a black hole with a finite horizon with AdS$_2\times S^2$
asymptotics.  For $N=6$, the solution takes a particularly simple form:
$F,G\sim [x_1^2+x_2^2+(x_3+iR)^2]^{-1/2}$. Namely, it is a complexified
version of $1/|{\bf x}-{\bf a}|$. The branch cut of the square root
leads to the multi-valuedness of the harmonic functions.  The second
example contains two stacks of codimension-2, circular branes with
non-Abelian monodromies. $\tau$ is not constant but depends on the
$\bbR^3$ coordinate~$\bf x$.  This solution was studied in
\cite{Fernandez-Melgarejo:2017dme} by a perturbative method, but we are
presenting the exact version.  This solution is free from CTCs but also
contains a continuous distribution of codimension-3 charge along the
symmetry axis of the configuration, which presumably underpins the
stability of the solution.  Because this charge density does not decay
rapidly enough at infinity, this solution cannot be regarded as a
microstate of a black hole.

In Appendix \ref{app:more_ex}, we discuss more 3D solutions obtained by
extending 2D solutions.  Unlike the solutions discussed above, they have
a puzzling feature; the 3D solution has a different monodromy structure
from the 2D solution that we started with.  This is unexpected because
extending the solution from 2D to 3D should not change the type of
codimension-2 branes which is characterized by the monodromy.  The
reason for this phenomenon and the physical relevance of these solutions
are unclear and deserve further investigation.

\bigskip
The organization of the paper is as follows.  In section
\ref{sec:multi-ctr_soln}, we briefly review the multi-center solutions
with sources of codimension two and three.  In section
\ref{sec:extend_to_3d}, we discuss how to start with a 2D harmonic
function $h(z)$ defined on a $z$-plane with non-trivial monodromies and
extend it to a 3D harmonic function $H({\bf x})$ defined in $\bbR^3$.
We present an extension formula that expresses~$H$ in terms of $h$ by a
kind of integral transform. In section \ref{sec:ex}, we apply the
extension formula to some 2D harmonic functions to derive exact 3D
harmonic functions with non-trivial monodromies, and discuss their
physical properties.  Section \ref{sec:disc} is devoted to discussion.
Appendices contain subjects not covered in the main text. Appendix
\ref{app:Legendre} explains Legendre functions $P_\nu,Q_\nu$ as well as
resonant Legendre functions $\cP_\nu,\cQ_\nu$ that are necessary in
constructing harmonic functions with nontrivial monodromies.  In
Appendix \ref{app:supertube}, we discuss how the ordinary supertube
solution fits in the framework of the current paper, where harmonic
functions are expanded in the basis of Legendre functions $P,Q$ and
their resonant versions $\cP,\cQ$.  In Appendix \ref{app:more_ex}, we
discuss more exact 3D solutions.

\section{Multi-center solutions with codimension 2 and 3}
\label{sec:multi-ctr_soln}

In this section we present a lightning review of the multi-center
solutions with sources of codimension two and three.  The main purpose
here is to establish notation. For more detail see \cite{Behrndt:1997ny,
Gauntlett:2002nw, Bates:2003vx, Bena:2004de, Gauntlett:2004qy,
Bena:2005va, Berglund:2005vb, Meessen:2006tu} (for solutions with
codimension-2 sources see \cite{Park:2015gka,
Fernandez-Melgarejo:2017dme}).

\subsection{Supersymmetric solutions}

Compactification of M-theory on a Calabi-Yau 3-fold $X$ leads to
ungauged 5D $\cN = 1$ supergravity with vector and hypermultiplets.  The
most general supersymmetric solution of this theory, with
hypermultiplets turned off and with a certain condition on $X$, were
classified in \cite{Gutowski:2004yv} (see also \cite{Gauntlett:2002nw,
Bena:2004de, Gutowski:2005id}).  When we apply this result to
$X=T^2_{45}\times T^2_{67}\times T^2_{89}$ and further assume a
tri-holomorphic $U(1)$ symmetry \cite{Gauntlett:2004qy}, we can reduce
the 11D solution to type IIA in ten dimensions, where the general
supersymmetric solution takes the following form:
\begin{align}
\label{njiw11Oct23}
\begin{split}
 ds_{\rm IIA,str}^2&=-{1\over \sqrt{V(Z-V\mu^2)}}(dt+\omega)^2
 +\sqrt{V(Z-V\mu^2)}dx_i dx_i\\
 &\qquad\qquad
 +
 \sqrt{Z-V\mu^2\over V}(Z_1^{-1}dx_{45}^2+Z_2^{-1}dx_{67}^2+Z_3^{-1}dx_{89}^2),\\
 e^{2\Phi}&={1\over  Z}\left({Z-V\mu^2\over V}\right)^{3/2},\qquad
 B_2=\left({K^I\over V}-{\mu\over Z_I}\right)J_I,
\end{split}
\end{align}
where $ds_{\rm IIA,str}^2$ is the 10D string-frame metric,
$dx_{45}^2\equiv dx_4^2+dx_5^2$ etc., $J_I\equiv dx_{2I+2}\wedge
dx_{2I+3}$, and $I=1,2,3$.  $x_i$ ($i=1,2,3$) are coordinates of a base
$\bbR^3$. There are also RR fields turned on; for their explicit
expressions see e.g.\ \cite[App.~E]{Park:2015gka} and
\cite{DallAgata:2010srl}.  Various functions appearing here are defined
in terms of the 3D harmonic functions $\cH=(V,K^I,L_I,M)$ (see
Eq.~\eqref{lozb5Dec23}) as
 \begin{align}
\begin{split}
 Z&=Z_1Z_2Z_3,\qquad
 Z_I=L_I+{1\over 2}C_{IJK}V^{-1}K^J K^K,\\
 \mu&=M+{1\over 2}V^{-1}K^IL_I+{1\over 6}C_{IJK}V^{-2}K^IK^JK^K,
\end{split} 
\end{align}
where $C_{IJK}=|\epsilon_{IJK}|$. The 1-form $\omega$ is found by solving
\begin{align}
  *d\omega&=
 VdM- MdV +{1\over 2}(K^IdL_I -L^I dK_I )
 =\ev{\cH,d\cH},
 \label{mpzn13May10}
\end{align}
where $*$ is the Hodge star for flat $\bbR^3$ with metric $\sum_{i=1}^3
dx_i^2$.  The skew product $\ev{\cH,\cH'}$ is defined by
\begin{align}
\ev{\cH,\cH'}\equiv VM'-MV'+{1\over 2}(K^I L'_I-L^I K'^I).
\end{align}
Applying $d*$ on \eqref{mpzn13May10} implies
the integrability condition \cite{Denef:2000nb}
(see also \cite{Bena:2005va})
\begin{align}
 0=\ev{\cH,\bigtriangleup \cH}.\label{jequ6Dec23}
\end{align}

The complexified K\"ahler modulus for $T^2_{89}$ is
\begin{align}
 \tau^3&=B_{89}+i\sqrt{{\rm vol}(T^2_{89})}
 =
 \left({K^3\over V}-{\mu\over Z_3}\right)
 +i{\sqrt{V(Z-V\mu^2)}\over Z_3 V},
\label{rny25Oct14}
\end{align}
where the radii of 456789 directions have been all set to $l_s=\sqrt{\alpha'}$.
The moduli $\tau^1,\tau^2$ for the tori
$T^2_{45},T^2_{67}$ are defined similarly.  
Because the volume of the
torus is a real positive number, we must have
\begin{align}
 \Im \tau^I>0.
\end{align}
In supergravity, $\tau^I$ live in the moduli space
$(SL(2,\bbR)/SO(2))^3$.  In string theory, this reduces to
$(SL(2,\bbZ)\backslash SL(2,\bbR)/SO(2))^3$ due to the $SL(2,\bbZ)^3$
duality symmetry that identifies different values of $\tau^I$ as
physically equivalent.

\subsection{Codimension-3 sources}

Being harmonic functions, $\cH$ can have codimension-3 (pointlike)
sources  as
\begin{align}
 \cH=h+\sum_{p=1}^N {\Gamma_p\over |{\bf x}-{\bf a}_p|}.
\label{jybh20Oct23}
\end{align}
The constant $h=(v_0,k_0^I,l^0_I,m_0)$ determines the value of the moduli
at infinity, while $\Gamma_p=(v_p,k_p^I,l_I^p,m_p)$ represents the
D-brane charge at position ${\bf x}={\bf a}_p$ \cite{Bates:2003vx}:
\begin{align}
 v_p\leftrightarrow \text{D6(456789)}~,
 \quad
 \begin{array}{l}
 k^1_p\leftrightarrow \text{D4(6789)}\\[.5ex]
 k^2_p\leftrightarrow \text{D4(4589)}\\[.5ex]
 k^3_p\leftrightarrow \text{D4(4567)}\\
 \end{array},
 \quad
 \begin{array}{l}
 l_1^p\leftrightarrow \text{D2(45)}\\[.5ex]
 l_2^p\leftrightarrow \text{D2(67)}\\[.5ex]
 l_3^p\leftrightarrow \text{D2(89)}\\
 \end{array},
 \quad
 m_p\leftrightarrow \text{D0}.
\label{singIIAbrn}
\end{align}
The position ${\bf a}_p$ of the sources is not arbitrary but must satisfy
the condition
\begin{align}
 \sum_{q(\neq p)}{\ev{\Gamma_p,\Gamma_q}\over |{\bf a}_p-{\bf a}_q|}=\ev{h,\Gamma_p}
\end{align}
that is obtained by requiring that $\delta$ function terms vanish in the
integrability condition~\eqref{jequ6Dec23}~\cite{Denef:2000nb,
Bena:2005va}.

\subsection{Duality action on harmonic functions}

The theory with $X=T^2_{45}\times T^2_{67}\times T^2_{89}$ has duality
group $SL(2,\bbZ)_1\times SL(2,\bbZ)_2\times SL(2,\bbZ)_3$, coming from
the $T$-duality for three individual $T^2$s.  The eight harmonic
functions in $\cH$ transform in the $({\bf 2},{\bf 2},{\bf 2})$
representation.  More concretely, under $SL(2,\bbZ)_3$ for example, each
of the combinations
\begin{align}
 \begin{pmatrix} K^3 \\ V \end{pmatrix},\quad
 \begin{pmatrix} 2M \\ -L_3 \end{pmatrix},\quad
 \begin{pmatrix} -L_1 \\ K^2 \end{pmatrix},\quad
 \begin{pmatrix} -L_2 \\ K^1 \end{pmatrix}\label{erem1Mar23}
\end{align}
transforms as a $\bf 2$.   Namely,
\begin{align}
 \begin{pmatrix} K^3 \\ V \end{pmatrix}
 \to 
 M
 \begin{pmatrix} K^3 \\ V \end{pmatrix}
\end{align}
where
\begin{align}
 M=\mqty( a & b \\ c & d ),
\qquad ad-bc=1,\qquad
 a,b,c,d\in\bbZ,
\label{fdmn12Oct23}
\end{align}
and likewise for other pairs in \eqref{erem1Mar23}.
Under this transformation, the modulus $\tau^3$ in \eqref{rny25Oct14} transforms as
\begin{align}
 \tau^3\to {a \tau^3+b\over c \tau^3+d}
\end{align}
while $\tau^{1,2}$ are invariant.
We can find transformations under $SL(2,\bbZ)_{1,2}$ by cyclically
permuting the indices in \eqref{erem1Mar23}.  In string theory, two
configurations related by this duality are different descriptions of the
same physics and identified.  Therefore, spacetimes patched together by
duality transformations are valid backgrounds; for example, we can
consider a defect curve $\cC$ around which spacetime is glued by a
duality transformation, which is the main topic of the current paper.

\subsection{Solutions  only with $\tau^3\equiv \tau$}

As already discussed in the introduction, we focus on the special class
of solutions (called SWIP solutions \cite{Bergshoeff:1996gg}) in which
the eight harmonic functions are determined in terms of two complex
harmonic functions, $F$ and $G$, as in~\eqref{mnrg5Dec23}.  In this
case, the complexified torus moduli are
\begin{align}
  \tau^1=\tau^2=i,\qquad \tau^3={F\over G}\equiv\tau.\label{fvnl1Jun17}
\end{align}
As mentioned in the introduction (and as can be easily seen using
\eqref{erem1Mar23}), the pair $\smqty(F\\ G)$ transforms as a $\bf 2$ of
$SL(2,\bbZ)_3$, and $\tau$ transforms as
\begin{align}
 \tau \to {a\tau+b\over c\tau+d}.
\end{align}
This is a useful subsector in which $SL(2,\bbZ)_{1,2}$ are switched off.
Note that this is not a solution in which branes are wrapping only the
third torus $T^2_{89}$ and the dual 4-cycle.  All types of branes in
\eqref{singIIAbrn} are present in a balanced way so that $T^2_{45},
T^2_{67}$ are kept trivial.

The 10D fields \eqref{njiw11Oct23} become
\begin{align}
\label{jumm27Oct23}
\begin{split}
 ds_{\rm IIA,str}^2&=-{1\over  |G|^2\Im(\tau)}(dt+\omega)^2
 +|G|^2  \Im(\tau) \,dx_{123}^2
 + dx_{45}^2+dx_{67}^2+\Im(\tau)\,dx_{89}^2,
\\[1ex]
 B_2
 &
 =\Re(\tau)\,dx^8\wedge dx^9
 ,
 \qquad
 e^{2\Phi}=\Im(\tau).
\end{split}
\end{align}
We clearly see that $\Im(\tau)>0$ is necessary for the solution to be
physically acceptable.  On the other hand, \eqref{mpzn13May10} becomes
\begin{align}
 *d\omega&=\Re(F d\Gb-G d\Fb)
 .\label{eiiw12Oct23}
\end{align}

In the presence of codimension-3 sources,
the harmonic functions $F,G$ can be written as
\begin{align}
 F=h_F+\sum_p {Q_F^p\over |{\bf x}-{\bf a}_p|},\qquad
 G=h_G+\sum_p {Q_G^p\over |{\bf x}-{\bf a}_p|},
\end{align}
where the complex quantities $(h_F,h_G)$ and $(Q^p_F,Q^p_G)$ are
combinations of the real quantities $h,\Gamma_p$
in \eqref{jybh20Oct23}.  We
will refer to $(Q^p_F,Q^p_G)$ as complex charges.  For generic charges,
the centers represent black holes, and the area entropy of the center at
${\bf x}={\bf a}_p$ is
\begin{align}
 S={\pi\left|\Im(Q_F^p \bar{Q}_G^p)\right|\over G_4},\label{jzur20Oct23}
\end{align}
where the 4D Newton constant is $G_4=g_s^2l_s^2/8$.

In passing, we mention that a codimension-2 center in the SWIP solutions
can describe either a 4-charge 1/8-BPS center or a 2-charge 1/4-BPS
center, but not a 3-charge 1/8-BPS center or a 1-charge 1/2-BPS center
\cite[Appendix D]{Fernandez-Melgarejo:2017dme}.  A 1-charge 1/2-BPS
center is said to be ``primitive'', describing a fundamental object in
string theory, and played an essential role in constructing microstate
geometries with codimension-3 centers \cite{Bena:2005va,
Berglund:2005vb}. On the other hand, codimension-2 centers in the SWIP
solutions can describe a fundamental object in string theory, namely a
supertube, which locally preserves 1/2 of supersymmetry
\cite{Bena:2011uw}.  Therefore, the SWIP solution is particularly suited
for constructing microstates made of codimension-2 centers.

\subsection{Codimension-2 solutions}

As mentioned in the introduction, we can have codimension-2 solutions,
in which the harmonic functions contain codimension-2 sources along a
curve $\cC\subset \bbR^3$ and have a duality monodromy around it as
\eqref{gnzz5Dec23}.

One simple example is the supertube, mentioned in the introduction, for
which D2(45) and D2(67) branes polarize into NS5($\cC$4567)-branes.  Let
us parametrize the closed curve $\cC$ by ${\bf x}={\bf F}(\lambda)$ with
$0\le \lambda\le 2\pi$, ${\bf F}(0)={\bf F}(2\pi)$. The numbers of
D2(45) and D2(67) branes are set to be equal.  Then the complex harmonic
functions and the modulus~$\tau$ are given by \cite{Park:2015gka}
\begin{align}
 \mqty(F\\G) = \mqty(h_F+k(\gamma + if) \\ 1),\qquad \tau=h_F+k(\gamma+if),
\end{align}
where $k\in\bbZ$ and the scalars $f,\gamma$ are defined by
\begin{align}
 f={1\over 4\pi}\int_0^{2\pi}{|\dot{\bf F}(\lambda)|\,d\lambda\over |{\bf x}-{\bf F}(\lambda)|},\qquad
 \alpha = {dx_i\over 4\pi}\int_0^{2\pi}{\dot{F}_i(\lambda)d\lambda\over |{\bf x}-{\bf F}(\lambda)|},\qquad
  d\alpha = * d\gamma.
\end{align}
The harmonicity of $f,\gamma$ is obvious from the definition.  The
integrability of the last equation, $d*d\alpha=0$, can be shown using
the fact that ${\bf F}(\lambda)$ is periodic.  Because $\gamma$ is
nothing but ${1\over 4\pi}$ times the solid angle subtended by $\cC$
measured from ${\bf x}$ \cite{Jackson:1998nia}, it shifts as $\gamma\to
\gamma+1$ as we go around $\cC$ (or, or course, this can be more
directly shown \cite{Park:2015gka}). Therefore, we have the monodromy
\begin{align}
 \mqty(F\label{neyl6Dec23}\\ G)\to 
 M\mqty(F\\ G),\qquad
 M=\mqty(1&k\\ 0&1).
\end{align}
From \eqref{jumm27Oct23}, we see that
$B_2=(\Re(h_F)+k\gamma)\,dx_8\wedge dx_9$.  This shifts by $k$ units as
we go around~$\cC$, which is the correct behavior of $B_2$ around
$k$ NS5($\cC$4567)-branes.

It is not so difficult to construct solutions with multiple
codimension-2 sources with the same type of monodromy \eqref{neyl6Dec23}
with different values of $k$, but having multiple codimension-2 sources
with non-commuting monodromies is much more challenging, which is the
goal of this paper.

In Appendix \ref{app:supertube}, we discuss this standard supertube
solution in detail in the framework the current paper when $\cC$ is a
circle.

\section{Extending harmonic functions to 3D}
\label{sec:extend_to_3d}

We would like to study codimension-2 solutions with circular, ring-like
sources.  In particular, we focus on axisymmetric solutions in which all
the rings are parallel to the $x_1$-$x_2$ plane and their centers are on
the $x_3$-axis.

\subsection{Toroidal coordinate system}
\label{ss:toroidal_coords}

We will initially be interested in situations where the radii of the
rings are all approximately~$R$ and their centers are all near the
origin, although we will relax this condition later.  Let the typical
inter-distance between the rings be $|L|\ll R$.

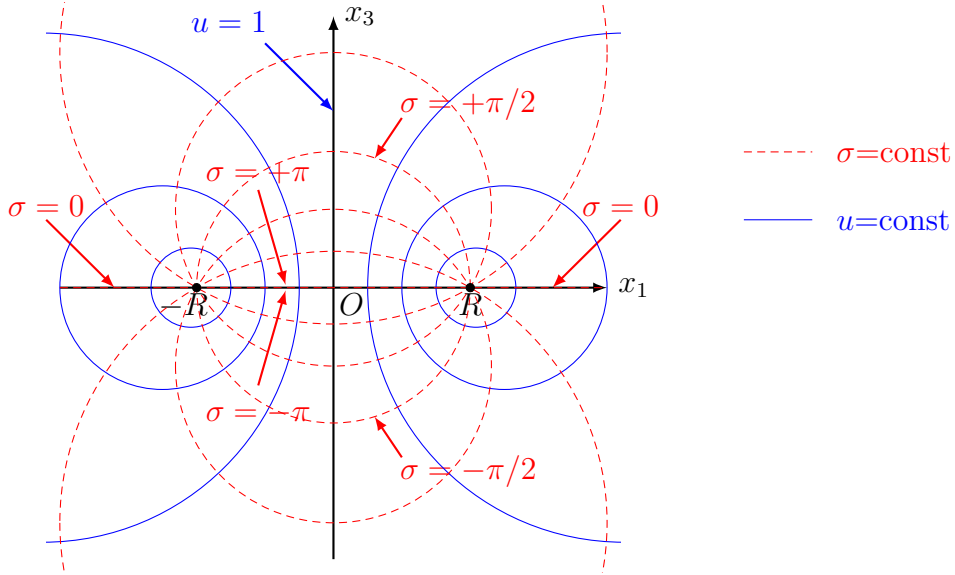
\begin{figure}[tb]
\begin{center}
 \begin{tikzpicture}[scale=1.8]
 \def\yoffset{0}
 \draw[densely dashed,color=red] (3.0,1) -- +(.5,0) node [right,xshift=5] {$\sigma$=const};
 \draw[color=blue] (3,0.5) -- +(.5,0) node [right,xshift=5] {$u$=const};
 %

 \draw[-latex,thick] (-2,0) -- (2,0) node [right] {$x_1$};
 \draw[-latex,thick] (0,-2) -- (0,2) node [right] {$x_3$};

 \draw[thick,latex-,color=blue] (0,1.3) -- +(-0.5,0.5) node [above left,xshift=5,yshift=0] {$u=1$};

 \draw[thick,color=red,latex-] ( 1.6,\yoffset) -- +( 0.5,0.5) node [above,yshift=-3] {$\sigma=0$};
 \draw[thick,color=red,latex-] (-1.6,\yoffset) -- +(-0.5,0.5) node [above,yshift=-3] {$\sigma=0$};

 \draw[thick,color=red,latex-] (-0.35,\yoffset+0.02) -- +(-0.2, 0.7) node [above,yshift=-2] {$\sigma=+\pi$};
 \draw[thick,color=red,latex-] (-0.35,\yoffset-0.02) -- +(-0.2,-0.7) node [below,yshift=-2] {$\sigma=-\pi$};

 \draw[thick,color=red,latex-] ( 0.30,\yoffset+.95) -- +(0.2, 0.3) node [above right,yshift=-5,xshift=-5] {$\sigma=+\pi/2$};
 \draw[thick,color=red,latex-] ( 0.30,\yoffset-.95) -- +(0.2,-0.3) node [below right,yshift= 5,xshift=-5] {$\sigma=-\pi/2$};


 \draw (0,0) node [below right,yshift=2,xshift=-2] {$O$};

 \draw[densely dotted] ( 1,\yoffset) -- ( 1,0) node [below,yshift=2] {$R$};
 \draw[densely dotted] (-1,\yoffset) -- (-1,0) node [below,yshift=2,xshift=-5] {$-R$};

 \clip (-2.1,-2.1) rectangle (2.1,2.1); 

 \foreach \u in {17/15, 5/3, 25/7} {
 \draw[color=blue] ({\u/sqrt(\u*\u-1)},\yoffset) circle ({1/sqrt(\u*\u-1)});
 \draw[color=blue] ({-\u/sqrt(\u*\u-1)},\yoffset) circle ({1/sqrt(\u*\u-1)});
 } 

   \foreach \sigma in {1/6,1/3,0.499,-1/3,-1/6} {
 \draw[densely dashed,color=red] (0,{1/tan(deg(\sigma*pi))+\yoffset}) circle ({1/sin(deg(\sigma*pi))});
 } 
 \draw[color=red,densely dashed] (-2,\yoffset) -- (2,\yoffset);

 \draw[fill=black] ( 1,\yoffset)circle (0.03);
 \draw[fill=black] (-1,\yoffset)circle (0.03);
 \end{tikzpicture}
\begin{quote}
 \caption{\label{fig:toroidal_coords}\sl Toroidal coordinates in the
 $x_2=0$ section. Blue solid lines represent constant-$u$ surfaces and
 red dotted lines represent constant-$\sigma$ surfaces.  As $u\to\infty$
 the constant-$u$ surface approaches the circle $x_1^2+x_2^2=R^2,x_3=0$,
 while as $u\to1$ the the constant-$u$ surface approaches the $x_3$-axis
 if $\sigma\neq 0$.  If $\sigma=0$, the $u\to 1$ limit corresponds to
 the 3D infinity.}
 \vspace*{-3ex}
\end{quote}
\end{center}
\end{figure}

In such situations,
it is useful to introduce toroidal coordinates $(u,\sigma,\psi)$; see
Figure~\ref{fig:toroidal_coords}.
The Cartesian coordinates
$(x_1,x_2,x_3)$ are related to the toroidal coordinates as
\begin{align}
x_1+ix_2=R\frac{\sqrt{u^2-1}}{u-\cos\sigma}e^{i\psi},\qquad 
 x_3=R\frac{\sin\sigma}{u-\cos\sigma}\,,\label{cartesian_to_toroidal}
\end{align}
where $u$ is a ``radial'' coordinate,
and $\sigma$ and $\psi$ are the angular variables, respectively, around and along 
the circle $\cC_0=\{(x_1,x_2,x_3)\,|\,x_1^2+x_2^2=R^2,x_3=0\}$.  The inverse relations are
\begin{align}
 u=\frac{\xv^2+R^2}{\Sigma}\,,\quad 
 \cos\sigma=\frac{\xv^2-R^2}{\Sigma}\,,\quad 
 \sin\sigma=\frac{2Rx_3}{\Sigma}\,,\quad 
 \tan\psi=\frac{x_2}{x_1}\,,\label{toroidal_ito_cartesian}
\end{align}
with
\begin{align}
\Sigma=\sqrt{(\xv^2-R^2)^2+4R^2 x_3^2}
 =|{\bf x}^2-R^2\pm 2i Rx_3|.\label{Sigma_def}
\end{align}
The flat 3D metric in the toroidal coordinates is
given by
\begin{align}
d\xv^2=
 \frac{R^2}{(u-\cos\sigma)^2}\left(\frac{du^2}{u^2-1}+d\sigma^2+(u^2-1)d\psi^2\right)\,.
 \label{3Dmetric_toroidal}
\end{align}
The range of $u$ is $[1,\infty]$.  $u=\infty$ corresponds to the circle
$\cC_0$, while $u=1$ corresponds to the $\bbR^3$
infinity (for $\sigma=0$) and the $x_3$-axis (for $\sigma\neq 0$).

The Laplace equation $\bigtriangleup H=0$ in the toroidal coordinate is
separable.  If we assume
\begin{align}
 H=\sqrt{u-\cos\sigma\,}\, e^{\pm i(\nu+\half)\sigma}h(u),
\label{iqhz11Dec23}
\end{align}
we find that $h$ must satisfy
\begin{align}
 \bigtriangleup H\propto 
 \left[(1-u^2)\p_u^2-2u\p_u+\nu(\nu+1)
 \right]h(u)=0,\label{ghns11Dec23}
\end{align}
which is the Legendre differential equation.  So, the general expression
for $H$ is
\begin{align}
 H(u,\sigma)=\sqrt{u-\cos\sigma\,}e^{\pm i(\nu+\half)\sigma}(AP_\nu(u)+BQ_\nu(u))
 \equiv H_0
 ,\label{gmyq11Dec23}
\end{align}
where $P_\nu(u)$ and $Q_\nu(u)$ are the Legendre function of the first
and second kinds, respectively.\footnote{There are different conventions
for $Q_\nu(u)$, specified by ``type''.  We use $Q_\nu(u)$ of type 3.
See Appendix \ref{app:Legendre} for detail.}  $P_\nu(u)$ is regular at
$u=1$ while $Q_\nu(u)$ goes like $-{1\over 2}\log(u-1)$ as $u\to 1$.
Unless $\nu\in\bbZ+1/2$, this $H$ changes by a phase as we go around $\cC_0$ and change
$\sigma\to\sigma+2\pi$.

We can also consider $H$ that has an additive monodromy. By setting
\begin{align}
 H=\sqrt{u-\cos\sigma\,}\, e^{\pm i(\nu+\half)\sigma}(
 \htilde(u)\pm i\sigma h(u)
 )
,
\end{align}
we find that $h(u)$ must satisfy \eqref{ghns11Dec23}, namely it is given
by \eqref{gmyq11Dec23}, while $\htilde(u)$ must satisfy the
Legendre equation with an inhomogeneous term,
\begin{align}
 \left[(1-u^2)\p_u^2-2u\p_u+\nu(\nu+1)
 \right]\htilde(u)=
 -(2\nu+1)\,h(u).
\end{align}
This is called the resonant Legendre differential equation
\cite{Backhouse:1986} and the solution is given by
\begin{align}
 \htilde(u)=A\cP_\nu(u)+B\cQ_\nu(u),
\end{align}
where $\cP_\nu(u)\equiv \p_\nu P_\nu(u),\cQ_\nu(u)\equiv\p_\nu Q_\nu(u)$
are ``resonant Legendre functions'' whose detail can be found in
Appendix~\ref{app:Legendre}.  The expression for $H$ is
\begin{align}
 H=\sqrt{u-\cos\sigma\,}\, e^{\pm i(\nu+\half)\sigma}
 \Bigl[
 A\bigl(\cP_\nu(u)\pm i\sigma P_\nu(u)\bigr)
 +B\bigl(\cQ_\nu(u)\pm i\sigma Q_\nu(u)\bigr)
 \Bigr].\label{jscc11Dec23}
\end{align}
As we change $\sigma\to \sigma+2\pi$, this changes as
\begin{align}
 H\to e^{\pm 2\pi i(\nu+\half)}(H\pm 2\pi i H_0).\label{jfbq11Dec23}
\end{align}


\subsection{2D limit}

Let us consider zooming in onto the region near
$\xv=(R,0,0)$. In this limit, rings near that point look like straight
lines along $x_2$.  This limit can be taken by sending $u \to +\infty$, $\psi\to 0$,
for which
\begin{align}
 x_1\approx R+{R\over u }\cos\sigma,\qquad 
 x_2\approx 0,\qquad
 x_3\approx {R\over u }\sin\sigma.\label{gxay19Apr17}
\end{align}
If we regard the $x_1$-$x_3$ plane as a complex $z$-plane, identifying
$(x_1,x_3)=(R,0)$ with the origin $z=0$, then we have the following
relation:
\begin{align}
z
 \approx 
 (x_1-R)+ix_3
 \approx {R\over u}e^{i\sigma}.\label{iepw1Mar23}
\end{align}
In this limit, the metric becomes that of flat $z$-plane
(ignoring $x_2$) and the Laplace equation reduces to $\p_z\p_{\zb}H=0$,
meaning that $H$ is a sum of holomorphic and anti-holomorphic functions.
We will refer to this limit as the {\it 2D limit}.

\subsection{An extension formula}

In the 2D limit, the 3D Laplace equation reduces to the 2D Laplace
equation, whose solution is $H_{\rm 2D}(z)=h(z)+\tilde{h}(\zb)$.  Given
such $H_{\rm 2D}(z)$, is it possible to extend it to a solution of the
3D Laplace equation that is defined in the entire $\bbR^3$?  The answer
to this problem can be found by a matching technique discussed in
\cite{Fernandez-Melgarejo:2017dme}, although a closed expression was not
presented there.

Assume that we have a 2D solution $H_{\rm 2D}=h(z)$, which we assume to
be holomorphic (generalization to include an anti-holomorphic part is
straightforward). If the 2D solution has no source at $z=\infty$, then
$h(z)$ must be expandable in $1/z$ as
\begin{align}
 h(z)=\sum_k  {a_{k}\over \zh^{k+\half}},\qquad \zh\equiv{z\over L},\label{iuml1Mar23}
\end{align}
where $k+1/2>0$ but $k$ is not necessarily an integer.  The constant $L$
determines the scale of the 2D solution and can be complex in
general:
\begin{align}
 L=|L|\,e^{il},\qquad l\in\bbR.
\end{align}
We can scale the size of the structure of the 2D solution by changing
$L$.
Then the expression for $H(u,\sigma)$ that
extends $h(z)$ to the entire $\bbR^3$ is
\begin{subequations} 
\label{extension_formula}
 \begin{align}
 H(u,\sigma)
 &=
 \sqrt{u -\cos\sigma\over Z}\sum_k 
 {2^{k} (k!)^2\over(2k)!}{ a_{k}\over Z^{k}}P_{k}(u)
 \label{iupl1Mar23}\\
 &=
 \sqrt{u -\cos\sigma\over Z}\sum_k 
 {(k!)^2\,a_{k}\over(2k)!}
 {1\over 2\pi i}\oint_C {dt\over t-u}\left[{t^2-1\over  (t-u)Z}\right]^{k},
 \label{ciwj10Mar23}
 \end{align}
\end{subequations}
where
\begin{align}
 Z\equiv {R\over L} e^{i\sigma}
\end{align}
and in the second equality we used the Schl\"afli integral
representation of the Legendre function,~\eqref{flxk12Jul23}.
We will denote a 2D solution by a lowercase letter and the corresponding
3D solution by a capital letter.  We will henceforth refer to
\eqref{extension_formula} as the extension formula.

We can readily check that the 3D harmonic function \eqref{iupl1Mar23}
reduces to the 2D expression~\eqref{iuml1Mar23} in the following sense.
We can go to the $z$-plane by considering $u\to \infty$, with the dictionary
\eqref{iepw1Mar23}
\begin{align}
 z\leftrightarrow {R\over u}e^{i\sigma}={LZ\over u},\qquad
 \text{or}\qquad
 \zh\leftrightarrow {Z\over u}.
\end{align}
Furthermore, let us simultaneously send ${L\over R}\to 0$ so that the
inter-distance between rings is much smaller than their radii.
Concretely, we send
\begin{align}
 Z,u\to \infty \qquad \text{with $\zh={Z\over u}$ fixed}.\label{scaling_lim}
\end{align}
Namely, we zoom in onto the region near $\xv=(R,0,0)$ and, at the same
rate, scale down the ring structure, so that we end up with a
non-trivial structure on the $z$-plane.  If we take this limit and use
the relation (see \eqref{lmnz11Dec23})
\begin{align}
 P_{k}(u)\to {(2k)!\over 2^{k}(k!)^2} u^{k}\qquad (u\to \infty),
\end{align}
it is easy to see that \eqref{iupl1Mar23} reduces to \eqref{iuml1Mar23}.
We will refer to the limit \eqref{scaling_lim} as the {\it scaling limit}.

We have not specified the integration contour $C$ in
\eqref{ciwj10Mar23}.  In the Schl\"afli integral  \eqref{flxk12Jul23},
the contour for $P_k(u)$ goes counterclockwise around the pole $t=u$ if
$k\in\bbZ$ and around the logarithmic branch cut $[1,u]$ if
$k\notin\bbZ$.  Because of the infinite sum, the integrand of the $t$
integral \eqref{ciwj10Mar23} may develop extra branch cuts.  In that
case, we may have to enlarge $C$ to enclose them.  The correct choice
for $C$ is determined by the requirement that it reproduce the 2D
harmonic function $h(z)$ in the scaling limit.

The above $H$ is not a unique 3D harmonic function that reduces to the
given 2D one in the scaling limit.  We can add to $a_k$ a number,
such as $(L/R)^n$ with $n>0$, which vanishes in the scaling limit.
Also, we could have included $Q_k(u)$ in \eqref{iupl1Mar23}, because it
vanishes as $u\to \infty$ for $k>-1$ and would not change the scaling
limit of $H$.  However, including $Q_k(u)$ leads to divergent $H$ in 3D,
because $Q_k(u)$ blows up logarithmically at $u=1$.

Another expression for $H(u,\sigma)$ is
\begin{align}
H(u,\sigma) =\sqrt{u -\cos\sigma \over Z}
  {1\over 2\pi i}\oint_C {dt\over t-u}
 \int_0^1 dy\,\,
 h'\!\left({Z(t-u)\over y(1-y)(t^2-1)}\right)
\label{vut10Mar23}
\end{align}
where
\begin{align}
 h'(\zh)
 \equiv
 \sum_k (2k+1) a_{k}\zh^{-k}
 =-2\zh^{3/2}{dh\over d\zh}.\label{bjha10Mar23}
\end{align}
The above expression can be derived using the identity
\begin{align}
 {(k!)^2\over(2k)!}
 &=
(2k+1)\int_0^1 dy\, y^k(1-y)^k.
\end{align}

Being a sum of harmonic functions, the function $H$ as given in
\eqref{iupl1Mar23} must be harmonic.  We can make the
harmonicity of $H$ manifest by going to  $w$ defined by
\begin{align}
 w= {t^2-1\over a(t-u)Z},\label{mgha6Nov23}
\end{align}
where $a$ is constant, so that
\begin{align}
 t={aZ\over 2} \left(w + \sqrt{w^2 - {4 u\over a Z}w +{4\over a^2Z^2}}\right),\qquad
 {dt\over t-u}={dw \over \sqrt{w^2 - {4 u\over a Z}w +{4\over a^2Z^2}}}.
\end{align}
The expression \eqref{ciwj10Mar23} now takes the form
\begin{align}
 \sqrt{u-\cos\sigma\over Z} \oint {dt \over t-u}\, \, \Phi\!\left({t^2-1\over a(t-u)Z}\right)
= \sqrt{u-\cos\sigma\over Z} \oint {\Phi(w)\,dw \over \sqrt{w^2 - {4 u\over a Z}w +{4\over a^2Z^2}}}\label{gzrp29Sep23} 
\end{align}
with a suitable function $\Phi$.  One can easily check that the
integrand in the last expression is harmonic for any~$w$.
Therefore, this is harmonic for an arbitrary function $\Phi(w)$, as long
as the specification of the contour $C$ does not depend on $u,\sigma$.
This arbitrariness corresponds to the arbitrary holomorphic function in
the 2D limit. 
Anti-holomorphic functions correspond to setting
$Z\to \bar{Z}$ in this expression.
The contour~$C$ in the $t$-plane can often be deformed in the $w$
coordinate into a contour encircling the branch cut $[w_-,w_+]$, where
$w_\pm$ are the roots of the radicand in \eqref{gzrp29Sep23}:
\begin{align}
 w^2-{4u\over aZ}w+{4\over a^2Z^2}=(w-w_+)(w-w_-),
\qquad w_\pm = {2\over aZ}(u\pm \sqrt{u^2-1}).
\label{jxky30Nov23}
\end{align}
For large positive $Z$ ($\gg u$), we have $0<w_-<w_+<1$.

\section{Examples}
\label{sec:ex}

Let us take some simple examples of 2D solutions $(f(z),g(z))$, apply
the extension formula~\eqref{extension_formula} to them to construct 3D
solutions $(F(u,\sigma),G(u,\sigma))$, and discuss their  properties.

\subsection{2D solutions from torus fibration}

We would like to start with some 2D harmonic functions $(f(z),g(z))$ as
a seed.  They are holomorphic functions which undergo nontrivial
$SL(2,\bbZ)$ monodromies as we go around branes which are pointlike on
the $z$-plane.  Moreover, their ratio $f/g=\tau$ must satisfy $\Im
\tau>0$.  The standard way to construct such $(f,g)$, used e.g.~in the
Seiberg-Witten theory \cite{Seiberg:1994rs, Seiberg:1994aj} and F-theory
\cite{Greene:1989ya, Vafa:1996xn}, is to consider fibration of a torus
over the $z$-plane and regard $(f,g)$ as the periods of the torus.

Concretely, consider a torus in the Weierstrass form:
\begin{align}
 y^2=x^3+p(z)x+q(z).\label{nkwf12Nov17}
\end{align}
where $p(z),q(z)$ are polynomials in $z$.
The modulus $\tau$ of the torus is known to be given by
\begin{align}
 j(\tau)={4(24p)^3\over 4p^3+27q^2}.\label{mooc28Mar23}
\end{align}
where $j(\tau)$ is Klein's $j$-invariant\footnote{Our convention
is such that $j(i)=24^3$, $j(e^{2\pi i/3})=0$.}
with a $q$-expansion
\begin{align}
 j(\tau)=2^3(q^{-1}+744+196884q+\cdots),\qquad q=e^{2\pi i \tau}.\label{ewrd4Oct23}
\end{align}
If we want
to control the value of $\tau$ at $z=\infty$, the degree of $p(z)$ and $q(z)$
must be $N/3$ and $N/2$ with some $N$.  For the degree of $p(z),q(z)$ to be
integral, $N$ must be an integer multiple of $6$.

The periods of the torus \eqref{nkwf12Nov17} are given by the integral of
the unique holomorphic 1-form $\lambda=dx/y$ along the $A$ and $B$ cycles:
\begin{align}
 f(z)&=\oint_A {dx\over\sqrt{x^3+p(z)x+q(z)}},\qquad
 g(z) =\oint_B {dx\over\sqrt{x^3+p(z)x+q(z)}}.
\label{cnn22Nov17}
\end{align}
For large $|z|$, by setting $x\to z^{N\over 6}x$, we can show that
these generically go like $z^{-{N\over 12}}$.  These periods give a
different expression for the modulus,
\begin{align}
 \tau(z)={f(z)\over g(z)}.
\end{align}

The discriminant of the cubic polynomial on the right hand side of
\eqref{nkwf12Nov17},
\begin{align}
 \Delta=4p^3+27q^2,
\end{align}
vanishes when some branch points collide and the torus becomes singular.
Indeed, from \eqref{mooc28Mar23} and \eqref{ewrd4Oct23}, we see that
$\Im\tau\to \infty$ at these points (unless $p=0$ there).  They give the
values of~$z$ at which branes are located.  Because $\Delta$ has degree
$N$, there are $N$ branes generically.  As we move in the $z$ space
around these branes, the cycles $A,B$ undergo non-trivial $SL(2,\bbZ)$
monodromies and so do the associated harmonic functions $f,g$.

\subsection{Constant-$\tau$ solutions}

\subsubsection{Constant-$\tau$ solutions in 2D}

Given the functions $p(z)$ and $q(z)$, equation \eqref{nkwf12Nov17}
describes fibration of a torus on the $z$-plane.  One particularly
simple case is when $\tau(z)$ is constant independent of $z$ (but the
harmonic functions $f(z),g(z)$ are generally nontrivial).  There are
three cases in which $\tau(z)$ is constant \cite{Sen:1996vd,
Dasgupta:1996ij}:
\begin{enumerate}[(i)]
 \item  \label{case:f=0}
 $p(z)=0$: \\
      In this case, we have $N\in 2\bbZ$ and 
\begin{gather}
j(\tau)=0,\quad \tau=e^{2\pi i/3},\qquad
 q\propto\prod_{i=1}^{N/2}(z-z_i),\qquad
 \Delta\propto
 \prod_{i=1}^{N/2}(z-z_i)^2.
\end{gather}	
The 2D harmonic functions \eqref{cnn22Nov17} are
\begin{align}
 g&= c\prod_i^{N/2}(z-z_i)^{-1/6},
\qquad
 f= e^{2\pi i/3}g,
\end{align}
where $c$ is constant.  There are 2 branes at each $z_i$.  Around each
stack of branes, there are an $SL(2,\bbZ)$ monodromy
$\smqty(\phantom{-}1&1\\-1&0)$, which has order 6, and a deficit angle
$\pi/3$.

 \item \label{case:g=0}
 $q(z)=0$: \\
      In this case, we have $N\in 3\bbZ$ and 
\begin{gather}
        j(\tau)=24^3,\quad \tau=i,\qquad
       p\propto \prod_{i=1}^{N/3}(z-z_i), \qquad \Delta\propto
 \prod_{i=1}^{N/3}(z-z_i)^3.
\end{gather}
The 2D harmonic functions are
\begin{align}
 g&= c\prod_i^{N/3}(z-z_i)^{-1/4},
\qquad
 f= ig,
\end{align}
where $c$ is constant. 
There are 3 branes at each $z_i$.  Around each stack of branes, there are
an $SL(2,\bbZ)$ monodromy 
       $\smqty(\phantom{-}0&1\\-1&0)$, which has order 4, and a deficit angle $\pi/2$.

 \item \label{case:f^3_g^2}
 $p(z)^3=\alpha^3q(z)^2$, $\alpha={\rm const}$:\\
      In this case, we have $N\in 6\bbZ$ and
\begin{align}
  j(\tau_0)&={4\cdot 24^3 \alpha^3 \over 4 \alpha^3+27},\quad
       q\propto\prod_{i=1}^{N/6}(z-z_i)^3,\quad
       p\propto\alpha\prod_{i=1}^{N/6}(z-z_i)^2,\quad
       \Delta\propto\prod_{i=1}^{N/6}(z-z_i)^6.\label{nfch3Oct23}
\end{align} 
The 2D harmonic functions are
\begin{align}
 g&= c\prod_i^{N/6}(z-z_i)^{-1/2},
\qquad
 f= \tau_0 g,\label{iaht12Dec23}
\end{align}
where $c$ is constant while $\tau_0$ is the constant determined by the
	first equation in \eqref{nfch3Oct23}.
There are 6 branes at each $z_i$.  Around each stack of branes, there are
an $SL(2,\bbZ)$ monodromy 
$\smqty(-1&\phantom{-}0\\ \phantom{-}0&-1)$, which has order 2, and a deficit angle $\pi$.
\end{enumerate}

\subsubsection{Constant-$\tau$ solutions with a single stack of branes}
\label{sss:single_stack}

Here we apply the extension formula \eqref{extension_formula} to the
case where there is a single stack of $N$ branes sitting at $z=0$.  
We take $N$ to be general, not specifying which of the cases
(i)--(iii) we are considering.  The 2D harmonic functions are
\begin{align}
 g=z^{-N/12},\qquad f=\tau_0 z^{-N/12},
\end{align}
where $\tau_0$ is the constant value of the torus modulus which
depends on the cases (i)--(iii).  $N$ can be an integer multiple of 2,
3, or 6, depending on the case (i), (ii) or (iii). There can be an
overall constant which we ignore at this point.

There being only one term in the $1/z$ expansion, the 3D harmonic
function can immediately be found using \eqref{extension_formula}:
\begin{align}
 G&={c\over R}\sqrt{u-\cos\sigma\,}\, e^{-i{N\over 12}\sigma}P_{{N\over 12}-{1\over 2}}(u)\notag\\
 &={c\over\sqrt{\Sigma}}\left({\xv^2-R^2-2iRx_3\over \Sigma}\right)^{N\over 12}P_{{N\over 12}-{1\over 2}}(u),\label{ermb2Jun23}
\end{align}
and 
\begin{align}
 F=\tau_0 G.
\end{align}
In the above,
\begin{align}
 c={2^{{N\over 12}}\Gamma({N\over 12}+{1\over 2})^2\over \Gamma({N\over 6})}
 \left({L\over R}\right)^{N\over 12}R,
\end{align}
but we can replace this $c$ by an arbitrary complex number, absorbing the
arbitrary overall complex number that we could have multiplied $f,g$ by.

The large $r=|\xv|$ behavior is
\begin{align}
 G= {Q_G\over r}+\cdots,\qquad 
 F= {Q_F\over r}+\cdots,\label{ghfy30Nov23}
\end{align}
where
\begin{align}
 Q_G=c,\qquad Q_F=c\tau_0.\label{kgic20Oct23}
\end{align}
Therefore, seen from far away, the ring looks like a codimension-3
center with complex charges $(Q_G,Q_F)$
and  a black-hole entropy \eqref{jzur20Oct23} which becomes
\begin{align}
 S={\pi |c|^2 \Im(\tau_0)\over G_4}.\label{fszl7Dec23}
\end{align}
So, this configuration can be regarded as a microstate of a black
hole. Note that there is no constant term in \eqref{ghfy30Nov23} and
hence this configuration is asymptotically AdS$_2\times S^2$.  This is
due to the non-trivial monodromy around the ring; a constant in the
harmonic function would be inconsistent with the monodromy.  Although
this solution must be locally 1/2-BPS, it is not a smooth geometry and
is therefore a microstate solution rather than a microstate geometry
according to the terminology of \cite{Bena:2013dka}.

Let us study the properties of the solution.  From \eqref{eiiw12Oct23},
we find
\begin{align}
*d\omega
 =- {|c|^2 N (\Im\tau_0)\over 6R^2}(u-\cos\sigma)\,P_{{N\over 12}-{1\over 2}}(u)^2\, d\sigma.
\end{align}
Integrating this equation, we obtain
\begin{align}
 \omega = {|c|^2 N \Im(\tau_0)\over 6R}\left(\int_1^u P_{{N\over 12}-{1\over 2}}(u)^2\, du\right)\, d\psi,
\end{align}
where we chose the constant of integration so that this vanishes at the
3D infinity ($u=1$).  Using the $u\to 1$ behavior of $P_\nu(u)$, we find
that the large $r$ behavior of $\omega$ is
\begin{align}
 \omega 
 = {|c|^2 N R\Im(\tau_0) \sin^2\theta\over 3r^2} d\psi+\cdots.
\end{align}
Because angular momentum $J$ is read off from the $\cO(1/r)$ term in
$\omega$, we see that $J=0$.  This is as it should be, because of the
AdS$_2$ asymptotics.  This is interesting in view of the claim
\cite{Sen:2009vz, Dabholkar:2010rm, Chowdhury:2015gbk} that genuine
black-hole microstates must have $J=0$.

By looking at the metric \eqref{jumm27Oct23}, we see that the $\psi$
direction potentially develops CTCs.  Using the above expression for
$\psi$, we find that the $\psi\psi$ component of the metric is
\begin{align}
 g_{\psi\psi}
& ={(\Im\tau)|c|^2\over (u-\cos\sigma)P_{{N\over 12}-{1\over 2}}(u)^2}
 \left[
 -\biggl({N\over 6}\biggr)^2 \Bigl(\int_1^u P_{{N\over 12}-{1\over 2}}(u)^2\Bigr)^2+(u^2-1)P_{{N\over 12}-{1\over 2}}(u)^4
 \right].\label{jfbt7Dec23}
\end{align}
Using the large $u$ behavior of $P_{{N\over 12}-{1\over 2}}(u)$, we can
show that $g_{\psi\psi}$ is non-negative everywhere for $N\le 6$.
However, if $N>6$, the $g_{\psi\psi}$ becomes negative near the ring and
the $\psi$ direction becomes a CTC\@.  So, only $N\le 6$ solutions are
physically acceptable.  This is unlike F-theory 7-branes, along which 
metric is regular.  This is because our branes are not
made of a single species of branes but are supertubes with lower charges
dissolved in the worldvolume.

Next, let us look at the geometry in the directions orthogonal to
$\psi$, namely $(u,\sigma)$ directions.  Near the ring, $u=\infty$, we
find that the $u,\sigma$ part of the metric goes like
\begin{align}
 ds^2
 &\sim 
 \tau_2
 |c|^2 R^2 \left({|L|\over R}\right)^{N/6}
 u^{{N\over 6}-{2}}
 \left({du^2 \over u^2}+d\sigma^2\right)
 \notag\\
 &\sim 
 |c|^2 R^2 \left({|L|\over R}\right)^{N/6}
 \rho^{-{N\over 6}}
 ({d\rho^2}+\rho^2d\sigma^2)
\end{align}
where $\rho\equiv 1/u$.  This is a conical deficit geometry with opening angle
$2\pi(1-{N\over 12})$. The distance to the position of the brane,
$\rho=0$ $(u=\infty)$, is
\begin{align}
 \sim \int_\epsilon \rho^{-N/12}d\rho
 = \left[{\rho^{1-N/12}\over 1-N/12}\right]_\epsilon
\end{align}
where $\epsilon\ll 1$ is a cutoff.  This is finite for $N<12$ but, for
$N>12$, the distance diverges. If $N=12$, the distance is
logarithmically divergent.  Recall that, in F-theory, 7-branes introduce
conical deficits in the 2-dimensional base space; if $N\le 12$, the base
is noncompact but, if $N\ge 12$, the base space becomes compact (for
$N=12$ the base becomes a half-infinite capped cylinder, while for
$N=24$ it becomes $S^2$).  This is showing up in our context as the
branes becoming unreachable from asymptotic infinity in the $N\ge 12$
case.

\subsubsection*{$\bullet$ $N=6$}

Let us discuss in some detail the $N=6$ case, where expressions are
simple and where there are no physical subtleties discussed above.  The
3D harmonic function \eqref{ermb2Jun23} is simply
\begin{align}
 G
={c\over R}\sqrt{u-\cos\sigma}\,e^{-i\sigma/2}
 &
 ={c\over \sqrt{x_1^2+x_2^2+(x_3+iR)^2}}.
\label{nhtl22Jun23}
\end{align}
We see the branch ``point'' at $x_1^2+x_2^2=R^2$, $x_3=0$ explicitly; we
get a minus sign as we go through the ring.  Flipping the sign of all
harmonic functions at once does not lead to issues like a wrong
signature of the metric, because the signs cancel in the combinations
that enter the metric.  Near this branch point, $G$ goes like
$u^{1/2}\sim 1/ \sqrt z$, as expected from $g=1/\sqrt{z}$.  That this is
harmonic is clear because this is $[({\bf x}-{\bf a})^2]^{-1/2}$ with
${\bf a}=(0,0,-iR)$.  This type of harmonic function appeared in
\cite{Behrndt:1997ny} in a different context.

One may wonder if we can put this brane at multiple places, so that
\begin{align}
 G\stackrel{?}{=}
{c\over \sqrt{x_1^2+x_2^2+(x_3+iR)^2}}
+{c'\over \sqrt{x_1^2+x_2^2+(x_3+iR')^2}},\label{fstk7Dec23}
\end{align}
which is certainly harmonic.  However, this is not physically allowed
because this screws up the monodromy.  Going around once around the
branes will flip the sign of only one term, whereas the $SL(2,\bbZ)$
transformation (which is $\smqty(-1&0\\ 0&-1)$ in this case) must act on
the entire~$G$.

The 1-form $\omega$ is given by
\begin{align}
 \omega
 ={|c|^2\Im(\tau_0)\over R}(u-1)d\psi
 ={|c|^2\Im(\tau_0)\over R}\left({r^2+R^2\over \Sigma}-1\right)d\psi,
 \label{fswy7Dec23}
\end{align}
which is well-behaved and vanishes at infinity.  One can show that the
Komar form, $\epsilon_{\alpha\beta\mu\nu}\nabla^\alpha \xi^\beta 
$ where $\xi=\partial_\psi$, has vanishing spatial components.
Therefore, these branes carry no angular momentum even locally; this is
quite different from ordinary supertubes, which carry angular momentum
by worldvolume fluxes.

The $\psi\psi$ component of the metric, \eqref{jfbt7Dec23}, is
\begin{align}
 g_{\psi\psi}=2(\Im\tau)\,|c|^2{u-1\over u-\cos\sigma}
\end{align}
which is positive everywhere.  The $\psi$ circle is finite even on the
ring ($u\to \infty$) which is unlike ordinary supertubes along which the
metric gets null \cite{Emparan:2001ux}.  This is because the current
solution is not an ordinary supertube but a limit in which multiple
supertubes with non-commuting monodromies collapse to a point 
\cite{Sen:1996vd, Dasgupta:1996ij}.

\begin{figure}[htb]
\begin{quote}
\begin{center}
   \begin{tikzpicture}[scale=0.8]
 \tikzset{
    partial ellipse/.style args={#1:#2:#3}{
        insert path={+ (#1:#3) arc (#1:#2:#3)}
    }
 }
 \draw[-latex,thick] (0,-5) -- (0,5) node [right] {$x_3$};
 \draw[-latex,thick] (-5,0) -- (5,0) node [right] {$x_1$};
 \draw[color=gray] (0,0) circle (4);
 \node () at (45:4.5) {$M^2_\infty$};
 \draw[fill=black] (2,0) circle (0.1);
 \draw[fill=black] (-2,0) circle (0.1);
 \draw[color=gray] (-1.83,0.1) -- (-0.1,0.1);
 \draw[color=gray] (0.1,0.1) -- (1.83,0.1);
 \draw[color=gray] (-1.83,-0.1) -- (-0.1,-0.1);
 \draw[color=gray] (0.1,-0.1) -- (1.83,-0.1);
 \draw[color=gray] (2,0) [partial ellipse=-150:150:.2 and 0.2];
 \draw[color=gray] (-2,0) [partial ellipse=30:330:.2 and 0.2];
  \node () at (0, 2.5) [right,xshift=-2] {$u=1$};
  \node () at (0,-2.5) [right,xshift=-2] {$u=1$};
  \node () at (0.25,+0.45) [right,xshift=-2] {$M_N^2$};
  \node () at (0.25,-0.45) [right,xshift=-2] {$M_S^2$};
  \draw [latex-] (-2.15, 0.15) -- +(-0.5, 0.5) node [above,xshift=-5,yshift=-2] {$M^2_{\rm tub}$};
  \draw [latex-] (2.15, 0.15) -- +(0.5, 0.5) node [right,xshift=-5,yshift=5] {$M^2_{\rm tub}$};
  \draw [latex-] (1.4, 0.1) -- +(0.25, 1) node [above,xshift=2,yshift=-2] {$\sigma\!=\!\pi$};
  \draw [latex-] (1.4,-0.1) -- +(0.25,-1) node [below,xshift=7,yshift=2] {$\sigma\!=\!-\pi$};
  \draw [latex-] (-0.05, 0.1) -- +(-0.75, 1) node [above left,yshift=-4,xshift=4] {$\p M^2_N$};
  \draw [latex-] (-0.05,-0.1) -- +(-0.75,-1) node [below left,yshift= 4,xshift=4] {$\p M^2_S$};
 \end{tikzpicture}
 \caption{\label{fig:Gauss_sfs} \sl Gaussian surface..}
\end{center}
\end{quote}
\end{figure}
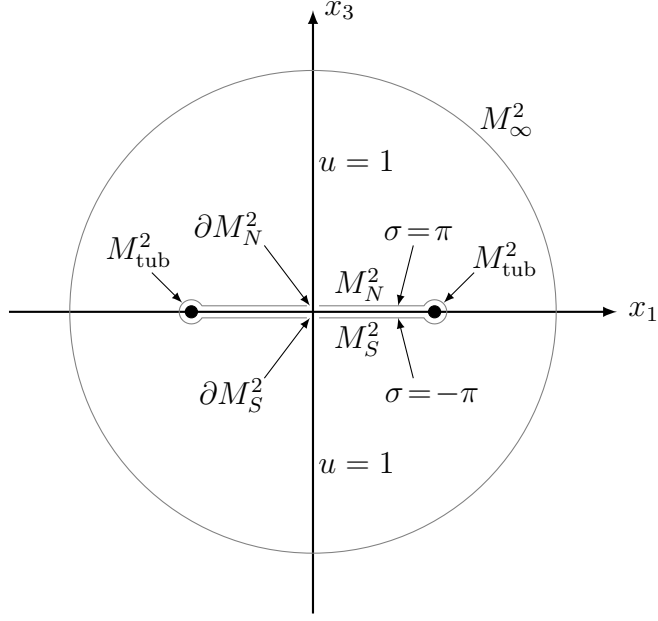

We know that the total charge is given by \eqref{kgic20Oct23}, but let
us study its distribution in some detail.  The charge can be measured by
\begin{align}
  Q_G=-{1\over 4\pi}\int _{M^2} *dG\label{efnl28Oct23}
\end{align}
where $M^2$ is a Gaussian surface enclosing the entire branes.  We can
take it to be $M^2_\infty$ in Figure~\ref{fig:Gauss_sfs}.  If $G$ were
single-valued, we could freely deform it as long as
singular sources are avoided.  However, in the present case where $G$ is
multi-valued, we cannot deform $M^2$ into a tubular surface enclosing
just the branes.  As in Figure \ref{fig:Gauss_sfs}, we can deform
$M^2_\infty$ into a tubular surface $M^2_{\rm tub}$ plus two disks,
$M^2_N$ and $M^2_S$, that are at $\sigma=\pi$ and $\sigma=-\pi$,
respectively, and touch the branes from inside.
In the present case,
\begin{align}
 *dG&=ce^{-i\sigma/2}{i(u-e^{-i\sigma})du+(u^2-1)d\sigma\over \sqrt{2}(u-\cos\sigma)^{3/2}}\wedge d\psi.
\end{align}
The contribution from  $M^2_{\rm tub}$ is
\begin{align}
 -{1\over 4\pi}\int_{M^2_{\rm tub}}*dG
 =
 {1\over 2} \int_{-\pi}^{\pi}  {ce^{-i\sigma/2}(u^2-1)\over 2(u-\cos\sigma)^{3/2}}d\sigma|_{u=u_c}=c\sqrt{2u_c},
\end{align}
where $u_c\gg 1$ is a regulator.  The contribution from $M^2_N$ and
$M^2_S$ is
\begin{align}
 -{1\over 4\pi}\int_{M^2_N+M^2_S}*dG
=
 {1\over 2} \int_1^{u_c} \left[ce^{-i\sigma/2}{i(u-e^{-i\sigma})\over 2(u-\cos\sigma)^{3/2}}du\right]^{\sigma=\pi}_{\sigma=-\pi}
 = c(1-\sqrt{2u_c}).
\end{align}
So, the Gaussian integration over the tubular region $M^2_{\rm tub}$ and
the disks $M^2_{N,S}$ individually diverge, but their sum is finite and
equal to \eqref{kgic20Oct23}. The same holds for $Q_F$.
The disk $M^2_{N,S}$ carrying charge by the multi-valuedness of fluxes
is an example of the so-called Cheshire charge that appears in the
presence of vortices with non-trivial monodromies called Alice strings
\cite{Schwarz:1982ec, Alford:1990mk, Preskill:1990bm} (for Alice strings
in string theory, see \cite{Harvey:2007ab, Okada:2014wma}).

Alternatively, we can compute the same Gaussian integral using the
1-form $\cG$ that satisfies
\begin{align}
 *dG=d\cG,
\end{align}
with which \eqref{efnl28Oct23} can be written as
\begin{align}
  Q_G=-{1\over 4\pi}\int _{\p M^2} \cG.
\end{align}
This way of computing the charge will be useful in the example
discussed later. In the present case, $\cG$ is given by
\begin{align}
 \cG &= {ic(u-e^{i\sigma})e^{-i\sigma/2}\over \sqrt{u-\cos\sigma}}d\psi.
\end{align}
This $\cG$ has Dirac strings on the $x_3$-axis.  So, as the boundary of
the Gaussian surface $M^2=M^2_{\rm tub}\cup M^2_N\cup M^2_S$, we can take small
circles around the $x_3$-axis at $\sigma=\pi$ ($\p M^2_N$) and at
$\sigma=-\pi$ ($\p M^2_S$).  At $u=1$,
\begin{align}
 \cG|_{u=1} 
&=c\, \sign(\sin\tfrac{\sigma}{2})\,
 d\psi.
\end{align}
Therefore, by adding the contributions from the north and south poles,
we obtain
\begin{align}
 Q_G=-{1\over 4\pi}(- 2\pi c-2\pi c)=c,
\end{align}
which is the same as \eqref{kgic20Oct23}.

We can  do a similar analysis for other values of $N$.

\subsection{Non-constant $\tau$ solution with two stacks of branes}
\label{ss:infSW}

As a more nontrivial example, consider the following torus:
\begin{align}
 y^2=(x^2-1)(x-\zh).\label{fxax4Oct23}
\end{align}
By shifting $x$, we can bring this curve into the Weierstrass form
\eqref{nkwf12Nov17}, with 
\begin{align}
   &\qquad p=-\left({\zh^2\over 3}+1\right),\quad q=-{2\over 27}\zh(\zh^2-9),
\label{isre23Nov17}
\end{align}
which corresponds to $N=6$.  Eq.~\eqref{mooc28Mar23} becomes
\begin{align}
 j(\tau)={f(z)\over g(z)}={ 512(\zh^2+3)^3\over (\zh^2-1)^2}.\label{fdhm23Oct23}
\end{align}
By looking at the values of $\zh$ at which the right-hand side diverges
and the order of the divergence, we see that there are two branes at
$\zh=1$ and another two branes at $\zh=-1$. These are points where the
discriminant~$\Delta$ vanishes.  In addition, the right-hand side
diverges at $\zh=\infty$, implying that there are two branes at
$\zh=\infty$.  In terms of $z=L\zh $, the branes are at $z=\pm
L,\infty$.

In \cite{Fernandez-Melgarejo:2017dme}, the 3D extension of the 2D
harmonic functions based on the the curve \eqref{fxax4Oct23} was studied
by a perturbative approach.  Here we use the extension formula
\eqref{extension_formula} to study the exact 3D extension.  The curve
\eqref{fxax4Oct23} is the Seiberg-Witten curve for pure $\cN=2$ $SU(2)$
Yang-Mills theory \cite{Seiberg:1994rs}, describing the vacuum moduli
space parametrized by the vev of the adjoint scalar. Here we are
borrowing the curve as a simple example of torus fibration that gives
non-trivial 2D harmonic functions.

\begin{figure}[htb]
\begin{center}
  \begin{tikzpicture}
 \tikzset{
    partial ellipse/.style args={#1:#2:#3}{
        insert path={+ (#1:#3) arc (#1:#2:#3)}
    }
 }
 \draw (-4.5,1) -- +(0,-0.4) -- +(-0.4,-0.4) node [xshift=6,yshift=7] {$x$};
  \node (mone) at (-3,0) [circle,fill=black,inner sep=0,minimum size=3pt,label=above:$\raisebox{1ex}{$-1~~$}$] {};
  \node (one) at (-1,0) [circle,fill=black,inner sep=0,minimum size=3pt,label=above:$\raisebox{1ex}{1}$] {};
  \node (z) at (1,0) [circle,fill=black,inner sep=0,minimum size=3pt,
   label=above:$\raisebox{1ex}{$\zh$}$] {};
  \node (inf) at (3,0) [circle,fill=black,inner sep=0,minimum size=3pt,label=above:$\raisebox{1ex}{$\infty$}$] {};
 \draw (mone) -- (one);
 \draw (z) -- (inf);
 \draw (-2,0) ellipse (1.2 and 0.5);
 \draw[-latex] (-2,-.5) -- +(0.01,0) node [below] {$A$};
  \draw[dashed] (0,0) [partial ellipse=0:180:1.2 and 0.5];
  \draw (0,0) [partial ellipse=180:360:1.2 and 0.5];
 \draw[-latex] (0,-0.5) -- +(-0.01,0) node [below] {$B$};
  end angle = 150];
  \end{tikzpicture}
 \caption{\label{fig:infSW_cycles_2d} \sl $A$ and $B$ cycles for the
 curve \eqref{fxax4Oct23}.}
\end{center}
\end{figure}
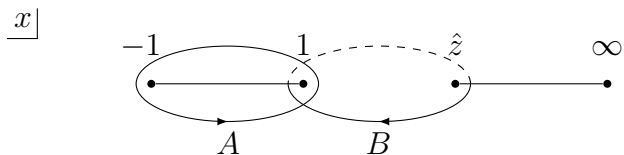
\begin{figure}[tb]
\begin{center}
  \begin{tikzpicture}[scale=1.0]
 \draw[fill=black] ( 1,0) circle (0.03) node [right] {$1$};
 \draw[fill=black] (-1,0) circle (0.03) node [below,xshift=4] {$-1$};
 \draw (-3.5,1.5) -- +(0,-0.4) -- +(-0.4,-0.4) node [xshift=6,yshift=7] {$\zh$};
\draw [thick,-latex] plot [smooth, tension=0.5] coordinates { (3,0) (1,0.5) (0.5,0) (1,-0.5) (3,-0.1)};
\draw [thick,-latex] plot [smooth, tension=0.5] coordinates { (3,-0.3) (0.8,-0.65) (-0.9,0.5) (-1.5,0) (-1,-0.5) (1,-.9) (3,-0.4)};
\draw [thick,-latex] plot [smooth, tension=1] coordinates { (3,0.1) (0,1.5) (-2.5,0) (0,-1.5) (3,-0.5)};
\node () at (1.0,0.75) {$M_1$};
\node () at (-0.9,0.75) {$M_{-1}$};
\node () at (-2.9,0) {$M_{\infty}$};
\draw[-latex] (0,1.5) -- +(-0.01,0);
\draw[-latex] (1,0.5) -- +(-0.01,0);
\draw[-latex] (-1.1,0.5) -- +(-0.01,0);
\end{tikzpicture}
\begin{quote}
 \caption{\label{fig:monod_paths}\sl Paths to go around branes and monodromies}
\end{quote}
\end{center}
\end{figure}
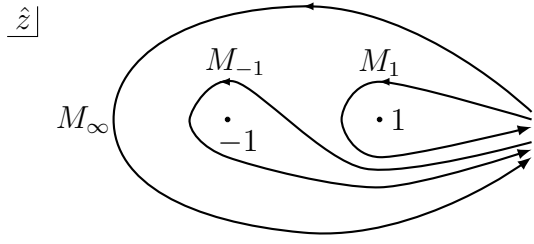

Specifically, let us take the branch cuts to be along $[-1,1]$ and
$[\zh,\infty]$, and take the A-cycle to go counterclockwise around the
cut $[-1,1]$ and the B-cycle to go clockwise around $[1,z]$. See Figure
\ref{fig:infSW_cycles_2d}. Then the 2D harmonic functions can be chosen
to be, with an arbitrary normalization,
\begin{subequations} 
 \begin{align}
 g(z)
 &={1\over\sqrt{2}\pi}\int_{-1}^{1} {dx\over \sqrt{(1-x^2)(\zh-x)}}
 ={\sqrt{2}\over\pi}{{\bf K}({2\over \zh+1})\over\sqrt{\zh+1}},
  \label{gjxa4Oct23}
 \\
 f(z)
 & ={i\over\sqrt{2}\pi}\int_{1}^{\zh} {dx\over \sqrt{(x^2-1)(\zh-x)}}
 ={i\sqrt{2}\over\pi}{{\bf K}(\tfrac{\zh-1}{\zh+1})\over \sqrt{\zh+1}}
  ,\label{ewrz23Oct23}
 \end{align}
\end{subequations}
where ${\bf K}(m)$ is the complete elliptic integral of the first
kind.\footnote{We take the Mathematica convention for the argument of
elliptic integrals.}  We could multiply $g,f$ by an overall factor.  The
monodromy of the cycles as we go counterclockwise around $\zh=1$, for
example, is
\begin{align}
 B\to B,\qquad A\to -2B+A,\label{glug27Oct23}
\end{align}
where the base point is taken to be large positive $\zh$.
Correspondingly, $f$ and $g$ transform as
\begin{align}
 f\to f,\qquad g\to -2f+g.\label{gltv27Oct23}
\end{align}
We represent this by the monodromy matrix
\begin{align}
 M_1=\begin{pmatrix}1&0\\ -2&1\end{pmatrix},\qquad
 M_{-1}=\begin{pmatrix}-1&2\\ -2&3\end{pmatrix},\qquad
 M_\infty=\begin{pmatrix}-1&2\\ 0&-1\end{pmatrix},
\label{gnxc23Oct23}
\end{align}
where we also presented the monodromy matrices for going counterclockwise around
$\zh=-1,\infty $.  These satisfy $M_1M_{-1}=M_\infty$.
See Figure \ref{fig:monod_paths}.

\begin{figure}[tb]
\begin{center}
  \begin{tikzpicture}
 \tikzset{
    partial ellipse/.style args={#1:#2:#3}{
        insert path={+ (#1:#3) arc (#1:#2:#3)}
    }
 }
 \draw (-4.5,1) -- +(0,-0.4) -- +(-0.4,-0.4) node [xshift=6,yshift=7] {$w$};
  \node (mone) at (-3,0) [circle,fill=black,inner sep=0,minimum size=3pt,label=above:$-1$] {};
  \node (wm) at (-1,0) [circle,fill=black,inner sep=0,minimum size=3pt,label=above:$\raisebox{1.5ex}{$w_-$}$] {};
  \node (wp) at (1,0) [circle,fill=black,inner sep=0,minimum size=3pt,label=above:$\raisebox{1.5ex}{$w_+$}$] {};
  \node (one) at (3,0) [circle,fill=black,inner sep=0,minimum size=3pt,label=above:$\raisebox{1ex}{$1$}$] {};
 \draw (mone) -- (wm);
 \draw (wp) -- (one);
 \draw (2,0) ellipse (1.2 and 0.5);
 \draw[-latex] (2,-.5) -- +(-0.01,0) node [below] {$B$};
  \draw[dashed] (0,0) [partial ellipse=0:180:1.2 and 0.5];
  \draw (0,0) [partial ellipse=180:360:1.2 and 0.5];
 \draw[-latex] (0,-0.5) -- +(0.01,0) node [below] {$A$};
  end angle = 150];
  \end{tikzpicture}
 \caption{\label{fig:infSW_cycles_w} 
 \sl Contours on the $w$-plane.}
\end{center}
\end{figure}
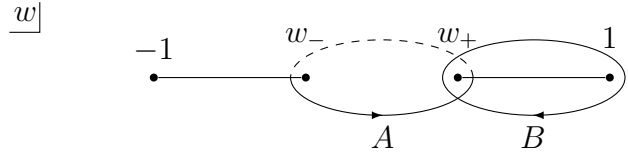

Let us find the 3D harmonic functions $F,G$ that correspond to $f,g$.
First, the large $z$ expansion of $g(z)$ is
\begin{align}
 g(z)={1\over\sqrt{2}}\sum_{k=0}^\infty
 {a_{2k}\over \zh^{2k+1/2}},\qquad
\end{align}
where
\begin{align}
 a_{2k}= {(4k)!\over 2^{6k}(2k)!(k!)^2},
 \qquad
 a_0=1,~ a_2={3\over 16},~ a_4={105\over 1024},~
 a_6={1155\over 16384},~
 \dots.
\label{kksz22Oct23}
\end{align}
This expansion can be derived by expanding the integrand of
\eqref{gjxa4Oct23} in $1/\zh$.  Applying the extension
formula~\eqref{ciwj10Mar23} to this gives the 3D harmonic function
 \begin{align}
\begin{split}
  G(u,\sigma)
 &=
 \sqrt{u -\cos\sigma\over 2Z}
 {1\over 2\pi i}\oint_C 
  \sum_{k=0}^\infty
 {(2k)!\over (k!)^2}
 {dt\over t-u}\left[{t^2-1\over  8(t-u)Z}\right]^{2k}
 \\
 &=
 {\sqrt{2(u -\cos\sigma)Z}\over \pi}
 \oint
 {dt\over \sqrt{(t^2-1)^2-16Z^2(t-u)^2}},
\end{split}
\label{fizm23Oct23}
\end{align}
which is again an elliptic integral.
Actually, it is more convenient to go to the $w$ coordinate defined in
\eqref{mgha6Nov23}.  In terms of $w$, \eqref{fizm23Oct23} can be written
as a still another elliptic integral as
\begin{align}
 G={1\over 2\sqrt{2}\pi}\sqrt{u-\cos\sigma\over Z}
 \oint_A{dw\over \sqrt{(w^2-1)(w^2-{u\over Z}w+{1\over 4Z^2})}}\label{mqkt6Nov23}
\end{align}
where we chose $a=4$. The contour $A$ is chosen as in Figure
\ref{fig:infSW_cycles_w}.  This choice of the contour is justified by
noting that \eqref{mqkt6Nov23} reduces to \eqref{gjxa4Oct23} in the
scaling limit; this is straightforward to show by setting $w=1/x$ and
taking the limit \eqref{scaling_lim}.
With this choice of the contour, the integral \eqref{mqkt6Nov23} can be
written  as
\begin{align}
 G
 &={1\over \sqrt{2}\pi}\sqrt{u-\cos\sigma\over Z}
 \int_{w_-}^{w_+}{dt\over\sqrt{(w^2-1)(w^2-{u\over Z}w+{1\over 4Z^2})}}
 \notag\\
 &={2\sqrt{2}\over \pi}
 \sqrt{(u-\cos\sigma)Z\over 4Z^2+4Z\sqrt{u^2-1\,}-1}
 \,\,
 {\bf K}\!\left({8Z\sqrt{u^2-1\,}\over 4Z^2+4Z\sqrt{u^2-1\,}-1}\right).
\label{bswv10Aug17}
\end{align}
By expanding this in powers of $1/Z$, one can check that this reproduces
the expansion \eqref{iupl1Mar23} with the Legendre function of the first
kind, $P_k(u)$, multiplying $Z^{-\half-k}$.  It is manifest that
this reduces to \eqref{gjxa4Oct23} in the scaling limit.

The location of branes can be determined by finding the values of
$(u,\sigma)$ at which some of the branch points collide.  This happens
when $w_+ = \pm 1$,\footnote{Here we assumed that $|Z|>1/2$.  If
$|Z|<1/2$, collision happens for $w_-=\pm 1$.} which gives
\begin{align}
(u,\sigma)=(u_*,l),~ (u_*,l-\pi),\qquad\qquad
 u_*\equiv{1\over 2}\left({2R\over |L|}+{|L|\over 2R}\right).
\label{def_u*}
\end{align}
In the scaling limit \eqref{scaling_lim} where $u,|Z|\to \infty$, this
reproduces the position of branes in the 2D solution, $\zh=\pm 1$ (or
$z=\pm L$).  As $|Z|={R\over |L|}$ decreases, $u_*$ decreases and the branes separate
away from each other other, until the singular limit $|Z|=1/2$ for which
$u_*=1$ and they are on the $x_3$-axis.  If we further decrease $|Z|$,
$u_*$ increases again and, in the $|Z|\to 0$ limit, the branes go back
to $u=\infty$.

The other 3D harmonic function $F$ cannot be obtained by the extension
formula \eqref{extension_formula}, because the large $\zh$ expansion of
the 2D harmonic function $f$ in \eqref{ewrz23Oct23} contains $\log \zh$:
\begin{align}
 f&={i\over \sqrt{2}\pi}
 \sum_{k=0}^\infty  {a_{2k}\log(8\zh)-b_{2k}\over \zh^{\half+2k}},\qquad
 b_0=0,~ b_2={5\over 16},~ b_4={389\over 2048}, ~ b_6={13327\over 98304},\quad\dots\label{ndlq24Oct23}
\end{align}
where $a_{2k}$ were defined \eqref{kksz22Oct23}.  One way to circumvent
this is to use the fact that, as we go around the brane at
$(u,\sigma)=(u_*,l)$, the doublet $(F,G)$ must undergo a monodromy by
the matrix $M_1$ in \eqref{gnxc23Oct23}.  Let $B$ be the 1-cycle going
clockwise around $[w_+,1]$ (see Figure \ref{fig:infSW_cycles_w}).  By
examining how the branch points move as go around
$(u,\sigma)=(u_*,l)$,
we can show that the 1-cycles $A,B$ transform as just as
\eqref{glug27Oct23}.  So, $F$ is given by
\begin{align}
  F(u,\sigma)
 &=
 {1\over 2\sqrt{2\,}\pi}\sqrt{u-\cos\sigma\over Z}
 \oint_{B}
 {dw\over \sqrt{(w^2-1)(w^2-{u\over Z}w+{1\over 4Z^2})}}
\notag
 \\
 &=
 {i\over \sqrt{2\,}\pi}\sqrt{u-\cos\sigma\over Z}
 \int_{w_+}^1
 {dt\over \sqrt{(1-w^2)(w^2-{u\over Z}w+{1\over 4Z^2})}}
\notag
 \\
 &=
 {2\sqrt{2\,}i\over \pi}
 \sqrt{(u-\cos\sigma)Z\over 4Z^2+4Z\sqrt{u^2-1\,}-1}
 \,\,
 {\bf K}\!\left({4Z^2-4Z\sqrt{u^2-1\,}-1\over 4Z^2+4Z\sqrt{u^2-1\,}-1}\right).
\label{erau7Nov23}
\end{align}
The large $Z$ expansion of this is
\begin{align}
 F&={i\over\sqrt{2}\pi}\sqrt{u-\cos\sigma}\sum_{k=0}^\infty
 \Biggl(
 {\log{8Z\over\sqrt{u^2-1}}\over Z^{1/2}}
 +{(3u^2-1)\log{8Z\over\sqrt{u^2-1}}-(5u^2-1)\over 16 Z^{5/2}}
 \notag\\
 &\qquad\qquad
 +{2(105u^4-90u^2+9)\log{8Z\over\sqrt{u^2-1}}-(389u^4-282u^2+21)\over 2048 Z^{9/2}}
+\cdots\Biggr).\label{ngcu24Oct23}
\end{align}
We can check that, in the scaling limit, this correctly reproduces the
expansion \eqref{ndlq24Oct23}.  
Or, more directly, we can see that the last expression of
\eqref{erau7Nov23} 
reduces to
\eqref{ewrz23Oct23} 
in the scaling limit.
One can confirm that, if we go around the
other stack of branes at $(u,\sigma)=(u_*,l-\pi)$, corresponding to the
$\zh=-1$ branes in the 2D limit, the 3D harmonic functions $(F,G)$
correctly transform according to the monodromy matrix $M_{-1}$ in
\eqref{gnxc23Oct23}. Therefore, the above $(F,G)$ are the correct 3D
extension of $(f,g)$.  Being the periods of a $w$-torus, it is
guaranteed that $\tau=F/G$ has $\Im \tau>0$.

Note that, unlike $G$, this $F$ diverges logarithmically at $u=1$,
namely on the $x_3$-axis.  This implies that this solution is not just
made of codimension-2 branes but also contains codimension-3 branes
continuously distributed along the $x_3$-axis.  This was not visible in
the perturbative analysis in the previous work
\cite{Fernandez-Melgarejo:2017dme}.  We can interpret this continuous
distribution of charge as manifestation of the brane at $\zh=\infty$ in
the 2D seed solution.  We will study the charge distribution in detail
below.

The fact that we have codimension-3 charge distributed on the $x_3$-axis
is presumably related to the stability of the solution.  In the 2D
setting, the position of the branes are free parameters because the
forces between branes cancel each other.  However, when we go to 3D, the
branes are bent and the balance of forces gets more nontrivial.  It is
possible that the charge sources on the $x_3$ is necessary for the stability of the two
circular branes.

In this solution, the source at infinity in the 2D seed is reappearing
in the 3D setting in a different guise.  This is unlike the ordinary
supertube, for which the problem at the 2D infinity is resolved in the
3D solution.  For the ordinary supertube, the 2D seed is $g=1,f={1\over
2\pi i}\log{z\over \mu}$.  This 2D solution does not make sense far away
from the center, $|z|>\mu$, because $\Im \tau = \Im (f/g)$ gets
negative.  However, the corresponding 3D solution is regular, even on
the $x_3$-axis, as discussed in Appendix \ref{app:supertube}\@.  We do
not have a good understanding of why the ordinary supertube and the
current example are different in this regard; we will discuss this
matter further in section \ref{sec:disc}.

\subsubsection{Charge distributions}

From the explicit expression \eqref{bswv10Aug17}, one can
straightforwardly show that the large $r$ behavior of $G$ is
\begin{align}
 G= {Q_G\over r}+\cdots,\qquad
 Q_G=\sqrt{LR \over 1-{L^2\over 4R^2}}.\label{hphf1Dec23}
\end{align}
$Q_G$ is the total $G$ charge of the system.  However, because of the
nontrivial monodromies, it is not clear whether we can think of this
charge as the sum of charges carried by the two separate stacks of
branes or not.  Moreover, the distribution of the charge for $F$ is more
non-trivial because of the aforementioned divergence on the
$x_3$-axis. Here we study in detail the distribution of charges for $G$
and $F$.  Because the analysis is slightly lengthy, we present a summary
at the end of this section on page \pageref{lmgu22Dec23}.

As
before, charges can be measured by integration over a Gaussian
surface $M^2$ as
\begin{align}
 Q_G=-{1\over 4\pi}\int_{M^2}*dG,\qquad
 Q_F=-{1\over 4\pi}\int_{M^2}*dF.
\end{align}
In the present case, $G,F$ are written as contour integrals on the
$w$-plane as
\begin{align}
 G=\oint_A \omega,\qquad
 F=\oint_B\omega,
\end{align}
where
\begin{align}
 \omega&={1\over 2\sqrt{2}\pi}\sqrt{u-\cos\sigma\over Z}
 {dw\over \sqrt{(w^2-1)(w^2-{u\over Z}w+{1\over 4Z^2})}}
\end{align}
is a 1-form on the $w$-plane and a scalar in $\bbR^3$.  Since $\omega$
is harmonic in $\bbR^3$ as we discussed around \eqref{gzrp29Sep23}, we
can define $\Omega$ by
\begin{align}
 *d\omega=d\Omega,
\end{align}
where $d$ is the exterior derivative with respect to
$\bbR^3$. 
$\Omega$ is a 1-form on
the $w$-plane and a 1-form in $\bbR^3$.
Explicitly, in the present case, $\Omega$ is given by
\begin{align}
 \Omega={iR\over 2\sqrt{2}\pi}{1\over \sqrt{(u-\cos\sigma)Z\,}}
 {[(u-e^{i\sigma})w+{1\over 2Z}(e^{i\sigma}-1)]dw
 \over (w-{L\over 2R})\sqrt{(w^2-1)(w^2-{u\over Z}w+{1\over 4Z^2})}}d\psi.\label{fdic16Nov23}
\end{align}
In terms of $\Omega$, we can write the charges as
\begin{align}
 Q_G=-{1\over 4\pi}\int_{\p M^2}\oint_A \Omega,\qquad
 Q_F=-{1\over 4\pi}\int_{\p M^2}\oint_B \Omega.
\label{jjkl26Nov23}
\end{align}
We would like to use this expression to study the charge distribution.
Although we wrote~\eqref{jjkl26Nov23} as contour integrals along $A$ and
$B$, actually, the relevant contours change as we move in
$\bbR^3$. Moreover, because $\Omega$ has an additional pole at $w=L/2R$
compared with $\omega$, we must specify whether or not we go around this
pole.  We will discuss these issues below.

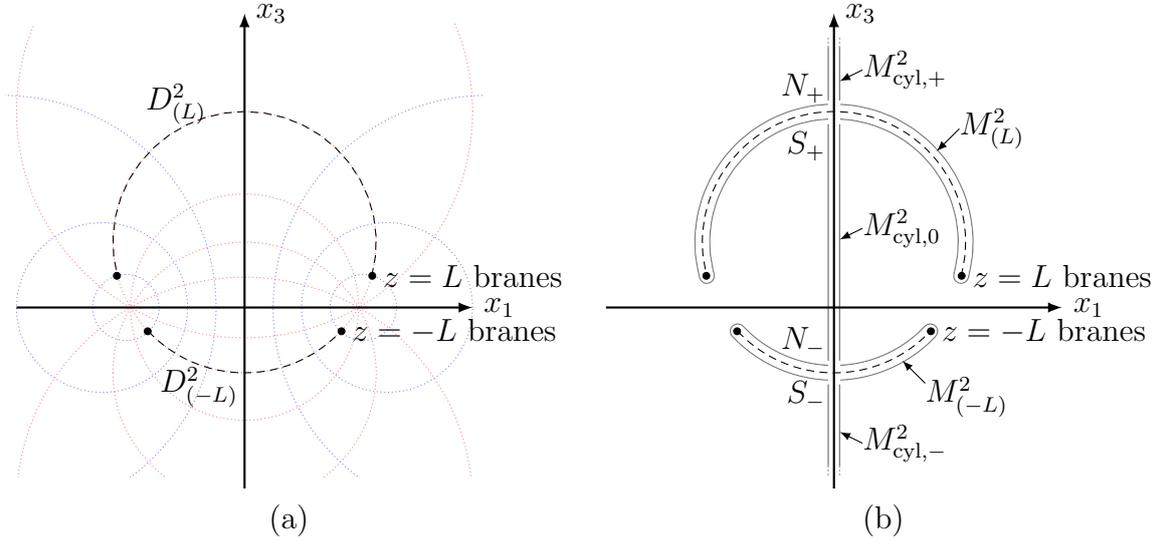
\begin{figure}[htb]
\begin{center}
\begin{tabular}{cc}
\begin{tikzpicture}[scale=1.5]
 \def\yoffset{0}
 \begin{scope}
  \clip (-2.1,-1.5) rectangle (2.1,2.5); 

 \foreach \u in {17/15, 5/3, 25/7} {
 \draw[color=blue!50!white,densely dotted] ({\u/sqrt(\u*\u-1)},\yoffset) circle ({1/sqrt(\u*\u-1)});
 \draw[color=blue!50!white,densely dotted] ({-\u/sqrt(\u*\u-1)},\yoffset) circle ({1/sqrt(\u*\u-1)});
 } 

  \foreach \sigma in {1/6,1/3,0.499,-1/3,-1/6} {
 \draw[densely dotted,color=red!50!white] (0,{1/tan(deg(\sigma*pi))}) circle ({1/sin(deg(\sigma*pi))});
 } 
 \end{scope}
 \draw[-latex,thick] (-2,0) -- (2,0) node [right] {$x_1$};
 \draw[-latex,thick] (0,-1.6) -- (0,2.6) node [right] {$x_3$};

 \draw [black,densely dashed,domain=-15:195] plot ({1.1547*cos(\x)}, {0.57735+1.1547*sin(\x)});
 \draw [black,densely dashed,domain=223:317] plot ({1.1547*cos(\x)}, {0.57735+1.1547*sin(\x)});

 \draw[fill=black] ( 1.12,0.28) circle (0.03) node [right] {$z=L$ branes};
 \draw[fill=black] (-1.12,0.28) circle (0.03);
 \draw[fill=black] ( 0.85,-0.21) circle (0.03)  node [right] {$z=-L$ branes};
 \draw[fill=black] (-0.85,-0.21) circle (0.03);


 \node () at (-0.6,1.8) {$D^2_{(L)}$};
 \node () at (-0.4,-0.7) {$D^2_{(-L)}$};
 \end{tikzpicture}
&
  \begin{tikzpicture}[scale=1.5]

 \draw [double=white,draw=gray,line width=0.5,double distance=5,domain=-15:195,line cap=round,smooth] plot ({1.1547*cos(\x)}, {0.57735+1.1547*sin(\x)});
 \fill[white] (-0.05,1.6) rectangle +(0.1,0.4);
 \draw [color=black!0!white,line width=5,domain=-15:195,line cap=round,smooth] plot ({1.1547*cos(\x)}, {0.57735+1.1547*sin(\x)});
 \draw [black,densely dashed,domain=-15:195,smooth] plot ({1.1547*cos(\x)}, {0.57735+1.1547*sin(\x)});
 \draw[latex-] (0.9,1.38) -- +(0.2,0.2) node [right,xshift=-4] {$M^2_{(L)}$};

 \draw [double=white,draw=gray,line width=0.5,double distance=5,domain=223:317,line cap=round,smooth] plot ({1.1547*cos(\x)}, {0.57735+1.1547*sin(\x)});
 \fill[white] (-0.05,-0.8) rectangle +(0.1,0.4);
 \draw [color=black!0!white,line width=5,domain=223:317,line cap=round,smooth] plot ({1.1547*cos(\x)}, {0.57735+1.1547*sin(\x)});
 \draw[latex-] (0.6,-0.47) -- +(0.2,-0.2) node [right,xshift=-4,yshift=-5] {$M^2_{(-L)}$};
 \draw [black,densely dashed,domain=223:317,smooth] plot ({1.1547*cos(\x)}, {0.57735+1.1547*sin(\x)});

 \draw[color=black!0!white,line width=5] (0,1.62) -- (0,-0.47);
 \draw[color=gray,line width=0.5] (-0.05,1.62) -- (-0.05,-0.47);
 \draw[color=gray,line width=0.5] (+0.05,1.62) -- (+0.05,-0.47);
 \draw[latex-] (0.05,0.6) -- +(0.2,0.1) node [right,xshift=-4] {$M^2_{\rm cyl,0}$};

 \draw[color=black!0!white,line width=5] (0,1.83) -- (0,2.4);
 \draw[color=gray,line width=0.5] (-0.05,1.83) -- (-0.05,2.3);
 \draw[color=gray,line width=0.5] (+0.05,1.83) -- (+0.05,2.3);
 \draw[color=gray,line width=0.5,densely dotted] (-0.05,2.3) -- (-0.05,2.4);
 \draw[color=gray,line width=0.5,densely dotted] (+0.05,2.3) -- (+0.05,2.4);
 \draw[latex-] (0.05,2.0) -- +(0.2,0.1) node [right,xshift=-4] {$M^2_{\rm cyl,+}$};

 \draw[color=black!0!white,line width=5] (0,-0.68) -- (0,-1.5);
 \draw[color=gray,line width=0.5] (-0.05,-0.68) -- (-0.05,-1.4);
 \draw[color=gray,line width=0.5] (+0.05,-0.68) -- (+0.05,-1.4);
 \draw[color=gray,line width=0.5,densely dotted] (-0.05,-1.4) -- (-0.05,-1.5);
 \draw[color=gray,line width=0.5,densely dotted] (+0.05,-1.4) -- (+0.05,-1.5);
 \draw[latex-] (0.05,-1.1) -- +(0.2,-0.1) node [right,xshift=-4] {$M^2_{\rm cyl,-}$};

 \draw[-latex,thick] (-2,0) -- (2,0) node [right] {$x_1$};
 \draw[-latex,thick] (0,-1.6) -- (0,2.6) node [right] {$x_3$};

 \draw[fill=black] ( 1.12,0.28) circle (0.03) node [right] {$z=L$ branes};
 \draw[fill=black] (-1.12,0.28) circle (0.03);
 \draw[fill=black] ( 0.85,-0.21) circle (0.03) node [right] {$z=-L$ branes};
 \draw[fill=black] (-0.85,-0.21) circle (0.03);

 \node () at (0.02,1.94) [left,xshift=0] {$N_+$};
 \node () at (0.02,1.43) [left,xshift=0] {$S_+$};
 \node () at (0.02,-0.30)[left,xshift=0] {$N_-$};
 \node () at (0.02,-0.78)[left,xshift=0] {$S_-$};

 \end{tikzpicture}
\\
(a) & (b)
\end{tabular}
\begin{quote}
 \caption{\label{fig:gauss_sfc}\sl (a) The two stacks of branes in 3D
 (black blobs).  The $x_2=0$ cross section is shown; the actual branes
 are extending along circles whose center is on the $x_3$-axis.  The
 stacks of branes that correspond to $z=L,-L$ branes in 2D are at $(u,\sigma)=(u_*,l),(u_*,l-\pi)$, respectively, in 3D in the toroidal
 coordinates. We take the branch cuts for the monodromies around them to
 be constant-$\sigma$ surfaces $D^2_{(\pm L)}$ shown by black dashed
 arcs.  We also show some constant-$\sigma$ (red) and constant-$u$
 (blue) curves by dotted lines.  (b) Gaussian surfaces.  We deform the
 Gaussian surface at infinity to the union $M^2_{\rm cyl,+}\cup
 M^2_{(L)}\cup M^2_{\rm cyl,0} \cup M^2_{(-L)}\cup M^2_{\rm cyl,-}$ that
 avoids singular sources and branch cuts. See the text for more
 detail. }
\end{quote}
\end{center}
\end{figure}

In view of the multi-valuedness of $F$ and $G$, we deform the Gaussian
surface $M^2$ as follows. Consider 2-surfaces $D^2_{(L)}$ and
$D^2_{(-L)}$ of disk topology, whose boundary is the branes at
$\sigma=l,l-\pi$, respectively.  We can choose $D^2_{(\pm L)}$
arbitrarily but, to be specific, we take them to lie on
constant-$\sigma$ surfaces (with $\sigma=l,l-\pi$). So,
\begin{align}
 D^2_{(L)}=\{(u,\sigma=l)\,|\,u\le u_*\},\qquad
 D^2_{(-L)}=\{(u,\sigma=l-\pi)\,|\,u\le u_*\};
\end{align}
see Figure \ref{fig:gauss_sfc}(a).  We take these to be the ``branch
cuts'' across which the values of $G,F$ jump.  We cannot deform $M^2$
past these branch cuts.  So, we deform $M^2$ into the union of the
following Gaussian surfaces:
\begin{align}
 M^2_{\rm cyl,+}
 \cup
 M^2_{(L)}
 \cup
 M^2_{\rm cyl,0}
 \cup
 M^2_{(-L)}
 \cup
 M^2_{\rm cyl,-}.
\label{ethl16Nov23}
\end{align}
Here, as shown in Figure \ref{fig:gauss_sfc}(b), $M^2_{\rm cyl,+}$ is a
cylinder of a very small radius enclosing the $x_3$-axis above
$D^2_{(L)}$, $M^2_{\rm cyl,-}$ is a cylinder below $D^2_{(-L)}$,
$M^2_{\rm cyl,0}$ is a cylinder between $D^2_{(L)}$ and $D^2_{(-L)}$,
and $M^2_{(\pm L)}$ is a surface that encloses $D^2_{(\pm L)}$.  The
boundaries of these surfaces are taken to be $\psi$-circles of a very
small radius going around the $x_3$-axis, along which Dirac strings are.
On these boundaries, $u=1$.  We will refer to the location of the
boundary of $M^2_{(L)}$ at its ``north pole'' (``south pole'') as $N_+$
($S_+$), and the location of the boundary of $M^2_{(-L)}$ at its ``north
pole'' (``south pole'') as $N_-$ ($S_-$), as in Figure
\ref{fig:gauss_sfc}(b).  Of course, where to take branch cuts is
physically irrelevant.

We want to compute the charges contained in each of the Gaussian
surfaces \eqref{ethl16Nov23}.  To evaluate the last expression in
\eqref{jjkl26Nov23}, we only need the value of $\Omega$ on the
boundaries of the Gaussian surfaces where $u=1$:
\begin{align}
 \Omega_\psi|_{u=1}
 &={\sqrt{LR}\over 2\pi}
 \sign(\sin\tfrac{\sigma}{2})
 {dw \over (w-{L\over 2R})\sqrt{w^2-1}}
 \equiv \Omegah
 .\label{euaw27Nov23}
\end{align}
We will be integrating this on the $w$-plane along a contour.  In the
$u\to 1$ limit, the cut $[w_+,w_-]$ around which contour $A$ goes (see
\eqref{jxky30Nov23}) collapses to a point $w={1\over 2Z}$ and
disappears.  If contour $A$ did not go around the pole at $w={L\over
2R}$, then we would get zero for $Q_G$, which is inconsistent with \eqref{hphf1Dec23}.  So, contour $A$ must go around it.  On the other hand,
in the $u\to 1$ limit, contour $B$, which is a closed path going around
$[w_-,1]$ is equivalent to twice the open contour going from $w={1\over
2Z}$ to $w=1$ (twice because we go on the first and second sheets).

\begin{figure}[tb]
\begin{center}
  \begin{tikzpicture}[scale=1.5]
 \draw [double=white,draw=gray,line width=0.5,double distance=5,domain=-15:195,line cap=round,smooth] plot ({1.1547*cos(\x)}, {0.57735+1.1547*sin(\x)});
 \fill[white] (-0.05,1.6) rectangle +(0.1,0.4);
 \draw [color=black!0!white,line width=5,domain=-15:195,line cap=round,smooth] plot ({1.1547*cos(\x)}, {0.57735+1.1547*sin(\x)});
 \draw [black,densely dashed,domain=-15:195,smooth] plot ({1.1547*cos(\x)}, {0.57735+1.1547*sin(\x)});

 \draw [double=white,draw=gray,line width=0.5,double distance=5,domain=223:317,line cap=round,smooth] plot ({1.1547*cos(\x)}, {0.57735+1.1547*sin(\x)});
 \fill[white] (-0.05,-0.8) rectangle +(0.1,0.4);
 \draw [color=black!0!white,line width=5,domain=223:317,line cap=round,smooth] plot ({1.1547*cos(\x)}, {0.57735+1.1547*sin(\x)});
 \draw [black,densely dashed,domain=223:317,smooth] plot ({1.1547*cos(\x)}, {0.57735+1.1547*sin(\x)});

 \draw[color=black!0!white,line width=5] (0,1.62) -- (0,-0.47);
 \draw[color=gray,line width=0.5] (-0.05,1.62) -- (-0.05,-0.47);
 \draw[color=gray,line width=0.5] (+0.05,1.62) -- (+0.05,-0.47);

 \draw[color=black!0!white,line width=5] (0,1.83) -- (0,2.4);
 \draw[color=gray,line width=0.5] (-0.05,1.83) -- (-0.05,2.3);
 \draw[color=gray,line width=0.5] (+0.05,1.83) -- (+0.05,2.3);
 \draw[color=gray,line width=0.5,densely dotted] (-0.05,2.3) -- (-0.05,2.4);
 \draw[color=gray,line width=0.5,densely dotted] (+0.05,2.3) -- (+0.05,2.4);

 \draw[color=black!0!white,line width=5] (0,-0.68) -- (0,-1.5);
 \draw[color=gray,line width=0.5] (-0.05,-0.68) -- (-0.05,-1.4);
 \draw[color=gray,line width=0.5] (+0.05,-0.68) -- (+0.05,-1.4);
 \draw[line width=0.5,densely dotted] (-0.05,-1.4) -- (-0.05,-1.5);
 \draw[line width=0.5,densely dotted] (+0.05,-1.4) -- (+0.05,-1.5);

 \draw[-latex,thick] (-2,0) -- (2.5,0) node [above] {$x_1$};
 \draw[-latex,thick] (0,-1.6) -- (0,2.6) node [right] {$x_3$};

 \draw[fill=black] ( 1.12,0.28) circle (0.03);
 \draw[fill=black] (-1.12,0.28) circle (0.03);
 \draw[fill=black] ( 0.85,-0.21) circle (0.03);
 \draw[fill=black] (-0.85,-0.21) circle (0.03);

\draw [thick,-latex] plot [smooth, tension=0.7] coordinates { (2.5,0.8) (1.5,2) (0.7,2.2) (0.05,1.8)};
\draw [thick,-latex] plot [smooth, tension=0.7] coordinates { (0.05,1.8) (0.7,1.8) (1.2,1.3) (1.4,0.7) (1.35,0.2) (1.1,0.1) (0.9,0.3) (0.85,0.8) (0.6,1.3) (0.05,1.7)};
\draw [thick,-latex] plot [smooth, tension=0.7] coordinates { (2.5,-0.4) (1.4,-0.9) (0.6,-1.0) (0.05,-0.65)};
\draw [thick,-latex] plot [smooth, tension=0.6] coordinates { (0.05,-0.65) (0.7,-0.65) (1.1,-0.3) (1.05,-0.1) (0.8,-0.05) (0.5,-0.15) (0.05,-0.5)};

\node () at (1.8,2) {$\cC_1$};
\node () at (1.55,1) {$\cC_2$};

\node () at (1.7,-1) {$\cC_3$};
\node () at (1.2,-0.5) {$\cC_4$};

 \node () at (0.02,1.94) [left,xshift=0] {$N_+$};
 \node () at (0.02,1.44) [left,xshift=0] {$S_+$};
 \node () at (0.02,-0.32)[left,xshift=0] {$N_-$};
 \node () at (0.021,-0.78)[left,xshift=0] {$S_-$};

 \end{tikzpicture}
\begin{quote}
 \caption{\label{fig:R3paths}\sl Paths in $\bbR^3$ to reach Gaussian
 surfaces}
\end{quote}
\end{center}
\end{figure}
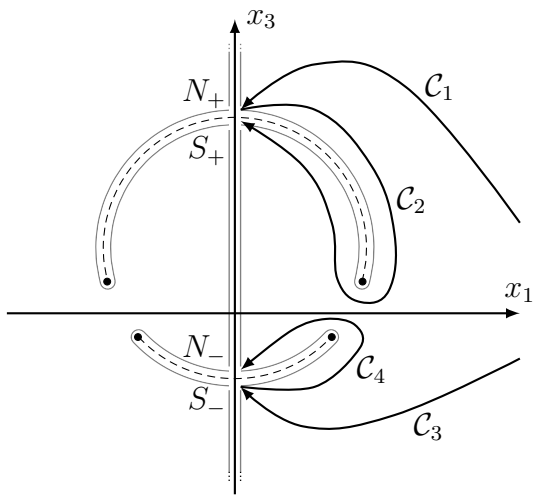

Let us start with the Gaussian surface $M^2_{\rm cyl,\pm}$.  We can reach
this region ($N_+$ and $S_-$) from the 3D infinity without changing the
contour defining $G,F$ (see the paths $\cC_1$ and $\cC_3$ in
Figure~\ref{fig:R3paths}).  Rather than the charge enclosed by the
entire $M^2_{\rm cyl,\pm}$, let us evaluate the charge density by
evaluating the charge in the infinitesimal interval $[\sigma,\sigma+d\sigma]$:
\begin{align}
 dQ_G&=-{1\over 2}\,d\sigma {\p \over \p \sigma} \oint_A\Omegah
\notag\\
&=-\sign(\sin\tfrac{\sigma}{2}) 
 {\sqrt{LR}\over 4\pi}\,d\sigma {\p \over \p \sigma} \oint_{w={L\over 2R}}
 {dw \over (w-{L\over 2R})\sqrt{w^2-1}}
 =0.\label{loez27Nov23}
\end{align}
This vanishes because nothing in the integral depends on $\sigma$.
Therefore, there is no $G$ charge on the $x_3$-axis, which we knew
because $G$ is regular there. For $F$, on the other hand, we get a
nonvanishing result because the lower bound of the integral depends on
$\sigma$:
\begin{align}
 dQ_F
 &=-{1\over 2}\,d\sigma {\p \over \p \sigma} \oint_B\Omegah
 \notag\\
 &=-\sign(\sin\tfrac{\sigma}{2})  {\sqrt{LR}\over 2\pi}\,d\sigma {\p \over \p \sigma} \int_{1\over 2Z}^1
 {dw \over (w-{L\over 2R})\sqrt{w^2-1}}
\notag\\
 &=
  {R\over 2\pi}
 {\sqrt{Z}\,d\sigma \over \left|\sin{\sigma\over 2}\right|\sqrt{1-4Z^2}}
={iR\,dx_3\over 2\pi\sqrt{(x_3+iR)^2-{L^2\over 4R^2}(x_3-iR)^2}},
\end{align}
where in the last expression we wrote the density per unit length in
$x_3$.  We see that on the $x_3$-axis there is a nonvanishing density
for $Q_F$ that goes like $\sim 1/x_3$.  Therefore, the total charge for
$F$ is not well defined in the sense that the total charge is
only conditionally convergent. 

Now we turn to $M^2_{(L)}$ that encloses the $z=L$ branes.  This
Gaussian surface has two boundaries at $N_+$ and $S_+$ (see Figure
\ref{fig:gauss_sfc}).  Considering the orientation of the boundaries,
the charge contained inside  $M^2_{(L)}$ is
\begin{align}
 Q_{G,F}|_{M^2_{(L)}}=
 -{1\over 2}\oint \Omegah|_{N_+}+\half \oint \Omegah|_{S_+}.
\end{align}
The contours in the first term is the same as those for $M^2_{\rm
cyl,+}$ but, in the second term, the contours change as we go from $N_+$
to $S_+$ along the path $\cC_2$ in Figure \ref{fig:R3paths}.  The change
in the contours is given by the monodromy matrix $M_1^{-1}=\smqty(1&0\\
2 & 1)$ where $M_1$ is given in \eqref{gnxc23Oct23}.  Therefore,
\begin{align}
 Q_G|_{M^2_{(L)}}
 &=
 -{1\over 2}\oint_A \Omegah|_{\sigma=l} 
 +{1\over 2}\oint_{2B+A} \Omegah|_{\sigma=l} 
 =
 \oint_{B} \Omegah|_{\sigma=l},
\label{neey27Nov23}
 \\
 Q_F|_{M^2_{(L)}}
 &=
 -{1\over 2}\oint_B \Omegah|_{\sigma=l} 
 +{1\over 2}\oint_{B} \Omegah|_{\sigma=l} 
 =
 0.
\end{align}
More explicitly,
\begin{align}
 Q_G|_{M^2_{(L)}}
 &=
 {\sqrt{LR}\over \pi}
 \int_{|L|\over 2R}^1
 {dw \over (w-{L\over 2R})\sqrt{w^2-1}}
.
\end{align}

The analysis of $M^2_{(-L)}$ that encloses the $z=-L$ branes is similar
to that of $M^2_{(L)}$.  It has two boundaries at $N_-$ and $S_-$ (see
Figure \ref{fig:gauss_sfc}).  The charge contained inside $M^2_{(-L)}$ is
\begin{align}
 Q_{G,F}|_{M^2_{(-L)}}=
 -{1\over 2}\oint \Omegah|_{N_-}+\half \oint \Omegah|_{S_-}.
\end{align}
The contours in the second term is the same as those for $M^2_{\rm
cyl,-}$ but, in the first term, the contours change as we go from $S_-$
to $N_-$ along the path $\cC_4$ in Figure \ref{fig:R3paths}.  The change
in the contours is given by the monodromy matrix $M_{-1}=\smqty(-1&2\\
-2 & 3)$ (see \eqref{gnxc23Oct23}).  Therefore,
\begin{align}
 Q_G|_{M^2_{(-L)}}
 &=
 -{1\over 2}\oint_{-2B+3A} \Omegah|_{\sigma=l-\pi} 
 +{1\over 2}\oint_{A} \Omegah|_{\sigma=l-\pi} 
 =
 \oint_{-A+B} \Omegah|_{\sigma=l-\pi},
\label{negg27Nov23}
 \\
 Q_F|_{M^2_{(-L)}}
 &=
 -{1\over 2}\oint_{-B+2A} \Omegah|_{\sigma=l-\pi} 
 +{1\over 2}\oint_{B} \Omegah|_{\sigma=l-\pi} 
 =
 \oint_{-A+B} \Omegah|_{\sigma=l-\pi}.
\end{align}

Finally let us consider $M^2_{\rm cyl,0}$.  We can reach it from 3D
infinity by following the paths $\cC_1\to\cC_2$ or $\cC_3\to\cC_4$ in
Figure \ref{fig:R3paths}.  If we follow $\cC_1\to \cC_2$, the charge
inside $M^2_{\rm cyl,0}$ is
\begin{align}
 Q_{F,G}|_{M^2_{\rm cyl,0}}
 &=-{1\over 2}\oint \Omegah|_{S_+}
 +{1\over 2}\oint \Omegah|_{N_-},
\end{align}
where the contours are just as for $M^2_{(L)}$. Therefore,
\begin{align}
 Q_G|_{M^2_{\rm cyl,0}}
 &
 =-{1\over 2}\oint_{2B+A} \Omegah|_{\sigma=l}
 +{1\over 2}\oint_{2B+A} \Omegah|_{\sigma=l+\pi}
 =-\oint_{B} \Omegah|_{\sigma=l}
 +\oint_{B} \Omegah|_{\sigma=l+\pi}
 ,\label{hnlt1Dec23}
 \\
 Q_F|_{M^2_{\rm cyl,0}}
 &=-{1\over 2}\oint_{B} \Omegah|_{\sigma=l}
 +{1\over 2}\oint_{B} \Omegah|_{\sigma=l+\pi},
\end{align}
where in the second equality of \eqref{hnlt1Dec23} we used the fact that
the $B$-cycle integral of $\Omegah$ does not depend on
$\sigma\in[l,l+\pi]$ if $0<l<2\pi$ (see \eqref{loez27Nov23}).  The
$B$-cycle integral does depend on~$\sigma$ and hence $Q_G|_{M^2_{\rm
cyl,0}}$ and $Q_F|_{M^2_{\rm cyl,0}}$ are nonvanishing.  Interestingly,
although $G$ has no charge source along the $x_3$-axis above or below
the branes ($M^2_{\rm cyl,\pm}$), it does have charge source in between
($M^2_{\rm cyl,0}$).\footnote{However, this depends on where to put the
branch cuts ($D^2_{(\pm L)}$).}

For $G$, there being no charge in $M^2_{\rm cyl,\pm}$, we can find the
total charge $Q_G$ by adding the contributions \eqref{neey27Nov23},
\eqref{negg27Nov23} and \eqref{hnlt1Dec23} as
\begin{align}
 Q_G
 &=
 Q_G|_{M^2_{(L)}}
 +Q_G|_{M^2_{(-L)}}
 +Q_G|_{M^2_{\rm cyl,0}}\notag\\
 &=\oint_B \Omegah|_{\sigma=l+\pi}
 +\oint_B \Omegah|_{\sigma=l-\pi}
 -\oint_A \Omegah|_{\sigma=l-\pi}.
\end{align}
As is clear from \eqref{euaw27Nov23},
$\Omegah|_{\sigma=l+\pi}+\Omegah|_{\sigma=l-\pi}=0$.  Therefore,
\begin{align}
 Q_G
 &=
 -\oint_A \Omegah|_{\sigma=l-\pi}\notag\\
 &={\sqrt{LR}\over 2\pi} \oint_{w={L\over 2R}} {dw\over(w-{L\over 2R})\sqrt{w^2-1}}
 =\sqrt{LR \over 1-{L^2\over 4R^2}},
\end{align}
which reproduces \eqref{hphf1Dec23} as it should.

\subsubsection*{Summary}

Let us summarize the computation of charges above.

The multi-valuedness of the fluxes means that the charges are not
carried only by the rings but also by the branch-cut disks whose
boundaries are the rings.  In addition, there is continuous distribution
of charges on the $x_3$-axis.  For the $G$ charge, there is no charge on
the $x_3$-axis above or below the branch-cut disks, but there is charge
on part of the $x_3$-axis between the branch-cut disks.  For the $F$
charge, there is charge distribution on the entire $x_3$-axis. Because
the $F$ charge density decays as $1/x_3$, the total $F$ charge is not
well-defined.  This means that, although this is a legitimate solution,
it is difficult to regard it as a microstate of a black hole with
definite charges.

Another observation is that the charges associated with individual rings
satisfy
\begin{align}
 Q_G|_{M^2_{(L)}}=0,\qquad
 Q_G|_{M^2_{(-L)}}= Q_F|_{M^2_{(-L)}}.
\label{lmgu22Dec23}
\end{align}
This is what was already observed in the perturbative analysis in
\cite{Fernandez-Melgarejo:2017dme}, and related to the fact that each
stack of branes is a 1/2-BPS supertube. Concretely, the $z=L$ and $z=-L$
branes must come from the supertube transition of 1/4-BPS codimension-3
centers of charge vectors $\Gamma_{L}=(a,(b,b,0),(0,0,a),0)$ and
$\Gamma_{-L}=(a',(b',b',-a'),(b',b',a'),{a'\over 2})$, respectively,
where $ab\neq 0,a'b'\neq 0$.  Because $\ev{\Gamma_{L},\Gamma_{-L}}
=aa'+bb'\neq 0$, this two-supertube configuration should be a bound
state.

\subsubsection{Absence of closed timelike curves}

Let us see whether the solution contains CTCs along $\psi$.  From
\eqref{jumm27Oct23} and \eqref{3Dmetric_toroidal}, the $\psi\psi$
component of the metric is proportional to
\begin{align}
 g_{\psi\psi}
 &\propto 
 -(u-\cos\sigma)^2\omega_\psi^2
 +R^2(\Im(F\Gb))^2(u^2-1).
\label{lqvg8Nov23}
\end{align}
The 1-form $\omega$ can be found by solving \eqref{eiiw12Oct23}.
Assuming that $\omega$ is independent of $\psi$, we find that 
$\omega_\sigma=\omega_u=0$ and that $\omega_\psi$ satisfies
\begin{align}
  \p_u \omega_\psi&=-{R\over u-\cos\sigma}\Re(F\p_\sigma \Gb-G\p_\sigma \Fb),
 \quad
 \p_\sigma \omega_\psi={R(u^2-1)\over u-\cos\sigma}
 \Re(F\p_u \Gb-G\p_u \Fb).
\label{eh9Nov23}
\end{align}

\bigskip First, let us look at the region near the $x_3$-axis where we
have a continuous distribution of charges, by setting $u=1+\epsilon$ and
expanding various quantities in $\epsilon$.  Using the explicit
expressions \eqref{bswv10Aug17}, \eqref{erau7Nov23}, we find
\begin{align}
 F\p_u \Gb-\Gb \p_u F
 ={iC\over \epsilon}
 +A+B\log \epsilon+\dots, \qquad
 C={Z_0(1-\cos\sigma)\over \pi|4e^{2i(\sigma - l)}Z_0^2-1|},
\label{lqkw8Nov23}
\end{align}
where $Z=Z_0e^{i(\sigma-l)}$, $Z_0={R\over |L|}$. $A$ and $B$ are some
complex quantities that depend on $\sigma$.  Therefore, the second
equation in \eqref{eh9Nov23} gives
\begin{align}
 \p_\sigma \omega_\psi 
 \sim \epsilon \Re( F\p_u \Gb-\Gb \p_u F)
 \sim A'\epsilon +B'\epsilon \log \epsilon+\text{(higher order in $\epsilon$)},
\label{edzw9Nov23}
\end{align}
where $A',B'$ are some real quantities that depend on $\sigma$.  Note
that the $\cO(1/\epsilon)$ term in \eqref{lqkw8Nov23} has dropped out.
Integrating this with respect to $\sigma$, we get
\begin{align}
 \omega_\psi 
 \sim \epsilon A''+B''\epsilon \log \epsilon+\cdots.
\label{fgzt9Nov23}
\end{align}
``$\,\cdots$'' here contain an integration constant which is independent
of $\sigma$ but can depend on $\epsilon$.  However, if we require that
$\omega$ vanish at 3D infinity $(\epsilon=0)$, then such a constant must
also vanish as $\epsilon\to 0$.  So, ``$\,\cdots$'' here are higher
order in $\epsilon$, or can be absorbed in $A'',B''$.  On the other
hand, we find
\begin{align}
 F\Gb=
 iC
 \log{1\over \epsilon} +\cdots.
\end{align}
Therefore,
\eqref{lqvg8Nov23} becomes
\begin{align}
 g_{\psi\psi}
 &\sim -\epsilon^2(A''+B''\log\epsilon)^2+ \epsilon  C^2 (\log\epsilon)^2,
\end{align}
which is positive for small $\epsilon\to 0$.  So, there is no CTC near
the $x_3$-axis.

\bigskip
Next, let us look at the region near the branes.
Let us focus on the region near the stack of branes at $(u,\sigma)=(u_*,l)$,
by setting
\begin{align}
 u=u_*+\Delta u,\quad \sigma=l+\Delta \sigma,\qquad u_*=Z_0+{1\over 4Z_0},
\end{align}
and expanding various quantities in $\Delta u$, $\Delta \sigma$.  We find
\begin{align}
 F\p_{ \sigma}\Gb-\Gb\p_{ \sigma}F
 ={1+4Z_0^2-4Z_0\cos l\over 4\pi (4Z_0\Delta u+i(4Z_0^2-1)\Delta \sigma)}+\cdots.
\end{align}
Plugging this into the first equation of
\eqref{eh9Nov23}, we find
\begin{align}
 \omega_\psi
 &=
 -{R(1+4Z_0^2-4Z_0\cos l)\over 16\pi Z_0(u_*-\cos l)} 
\log r_\perp+\cdots,\label{cgc9Nov23}
\end{align}
where
\begin{align}
r_\perp = {R\over u_*-\cos l}
 \sqrt{{(\Delta u)^2\over u_*^2-1}+(\Delta \sigma)^2}
\end{align}
is the distance from the location of the brane.
On the other hand,
\begin{align}
 \Im(F\Gb)=
 {1+4Z_0^2-4Z_0\cos l\over 4\pi(4Z_0^2-1)}
 \log r_\perp+\cdots.
\label{cgg9Nov23}
\end{align}
If we plug 
\eqref{cgc9Nov23} and
\eqref{cgg9Nov23} into 
\eqref{lqvg8Nov23}, we find that the two contributions cancel:
\begin{align}
 g_{\psi\psi}=0\cdot
(\log r_\perp)^2
+\cdots.
\label{cse9Nov23}
\end{align}
So, the $\psi$ direction is null at the leading order.  This is
consistent with the fact that, up to U-duality, each stack of branes is
just an ordinary supertube along which the metric becomes null
\cite{Emparan:2001ux}.  The subleading terms in \eqref{cse9Nov23} cannot
be determined in this local analysis because we do not know the constant
of integration in \eqref{cgc9Nov23}, which is determined by the boundary
condition at the other end of space, that $\omega=0$ at $u=1$.  However,
physically we expect that there is no CTC much as for ordinary
supertubes.

%
%

\section{Discussions}
\label{sec:disc}

In this paper, we constructed examples of exact solutions that represent
non-Abelian supertubes in the framework of the multi-center solutions.
Multi-center solutions are characterized by 3D harmonic functions, and
supertubes correspond to codimension-2 singularities in the harmonic
functions.  In order to construct desired 3D harmonic functions, we
started with 2D harmonic functions that represent F-theory--like
configurations of branes in a 2D $z$-plane.  We extended those 2D
harmonic functions to 3D harmonic functions, by an extension formula,
which essentially replaces $z^{-k}$ in 2D by a Legendre function
$P_{k-\half}(u)$.

We worked out two explicit examples in the main text.  In the first
example, we took as a 2D seed the constant-$\tau$ solution with a single
stack of branes.  The resulting 3D solution contains a single circular
stack of codimension-2 branes, around which there is a non-trivial
monodromy.  If the number of branes is $N\le 6$, the solution is
free from physical issues such as CTCs and can be thought of as a new
microstate of a black hole with a finite entropy in AdS$_2\times S^2$.
This example is perhaps a little too simple having only one ring.  As a
next example, we considered a solution that has two stacks of circular
branes with non-Abelian monodromies.  This solution has no CTCs and has
a curious feature that there is continuous distribution of charges on
the $x_3$-axis.  We gave a detailed analysis of the charges in this
configuration, and found that the charge density on the $x_3$-axis makes
it difficult to interpret this solution as a microstate of a black hole.
This extra charge presumably necessary for the stability of the
configuration.
In Appendix \ref{app:more_ex}, we considered more examples, but they
have a serious problem of the monodromy in the 3D solution being different
from that in the 2D solution that we started with.

Thus we have made important steps toward constructing general
black-hole microstates involving codimension-2 centers.  However, the
above issues must be clarified in order to construct more general
solutions.

Let us discuss points potentially relevant to these issues.  We used the
extension formula \eqref{extension_formula} to extend 2D seed solutions
to 3D solutions, but this procedure leaves some ambiguities.  The
extension formula essentially replaces $z^{-k}$ in 2D by the Legendre
function $P_{k-\half}(u)$.  However, we are free to include the other
Legendre function $Q_{k-\half}(u)$, because it does not spoil the
matching. Furthermore, there is freedom to change the coefficient of
$z^{-k}$ by a number that vanishes in the scaling limit.  To appreciate
these points, let us look at the 3D harmonic function $F$ that we
studied in the two-stack example in section~\ref{ss:infSW}.  We applied
the extension formula to the 2D harmonic function $g$ to construct the
3D harmonic function $G$, while $F$ was obtained from $G$ by the
monodromy requirement, without using the extension formula.  The large
$Z$ expansion of $F$ \eqref{ngcu24Oct23} can be written as
\begin{align}
 F&={i\over \sqrt{2}\pi}\sqrt{u-\cos\sigma}\, f,\qquad
f=f_{-1/2}Z^{-1/2}+ f_{-5/2}Z^{-5/2} + \cdots,
\end{align}
where
\begin{align}
\begin{split}
  f_{-1/2}
 &=
 -(\cP_0-i\sigma P_0)
 +P_0\log{4R\over L}
 +Q_0,
 \\
 f_{-5/2}
 &=
 {1\over 8}
 \left[-(\cP_2-i\sigma P_2)+\log{4R\over L}P_2+Q_2\right]
 -{1\over 16}P_2,
 \\
 f_{-9/2}&=
 {3\over 128}\left[-(\cP_4-i\sigma P_4)+\log{4R\over L}P_4  +Q_4\right]
 -{7\over 512}P_4,\qquad \dots
\end{split}
\label{nijz24Dec23}
\end{align}
We saw that, in the scaling limit, this $F$ reproduces the 2D harmonic
function $f$ \eqref{ndlq24Oct23}.  However,  \eqref{nijz24Dec23} also contains $Q_k(u)$,
which cannot be obtained by the extension formula.
We have another ambiguity in constructing 3D harmonic functions from 2D
ones.  Because the 2D seed is holomorphic, we assumed that the 3D
harmonic functions are also ``holomorphic'' in the sense discussed below
\eqref{gzrp29Sep23} or \eqref{mtgd22Dec23}.  However, as far as matching
is concerned, we could have also included ``antiholomorphic'' pieces as long
as they do not contribute in the scaling limit.  An interesting example
in this regard is the ordinary supertube solution discussed in
Appendix~\ref{app:supertube}.  In that solution, the harmonic functions
involve only $Q,\cQ$ and not $P,\cP$.  Furthermore, that harmonic
functions do not just have holomorphic parts but also antiholomorphic
pieces that do not contribute in the scaling limit.  The structure of
the harmonic functions, including the antiholomorphic part, is determined
by the monodromy requirement.
These examples may mean that, although there are ambiguities in
constructing 3D harmonic functions, the requirement of desired
monodromies is strong enough to completely fix those ambiguities.
This matter certainly deserves further investigation.  In
particular, it is important to clarify whether or not the charge distribution
on the $x_3$-axis of the two-stack example is an unavoidable consequence
of the monodromy requirement.

It is possible that the class of solutions we studied in the current
paper is too restrictive.  We focused on the so-called SWIP solutions
where the moduli $\tau_1,\tau_2$ are fixed to $i$.  This is not a very
mild assumption; for example, it excludes 1/2-BPS primitive centers of
codimension~3, which are crucial for the construction of the bubbling
microstate geometries \cite{Bena:2005va, Berglund:2005vb}.  It is
desirable to study more general solutions with one or both of $\tau_1$
and $\tau_2$ are allowed to vary.  Allowing one to vary may be
manageable, because the eight harmonic functions can be written in terms
of two pairs of complex functions each of which transforms under
$SL(2,\bbZ)$ \cite[Appendix C]{Fernandez-Melgarejo:2017dme}.

Another direction is to include both codimension-2 centers and
codimension-3 centers from the start \cite{Park:2015gka}.  In
particular, our second example suggests that having codimension-2
centers with non-commuting monodromies is not generally stable on its
own.  By explicitly including co\-di\-men\-sion-3 centers, we can hope
to find simple solutions representing genuine microstates of black
holes.  As mentioned above, for the additional codimension-3 center to
be 1/2-BPS primitive, we must go out of the SWIP class of solutions.

\section*{Acknowledgments}

We would like to thank Iosif Bena and Kazumi Okuyama for discussions.
We thank Jos\'e J. Fern\'andez-Melgarejo and Minkyu Park, and Hitoshi Sakai
for collaboration in the early stages of the project.  The work of MS
was supported in part by MEXT KAKENHI Grant Numbers 21K03552 and
21H0518.

\appendix

\section{Legendre functions and resonant Legendre functions}
\label{app:Legendre}

\subsection{Legendre functions}

The Legendre functions
are solutions of the Legendre differential equation
\begin{align}
 (1-z^2)f''-2z f'+\nu (\nu+1)f=0, \qquad \nu\in\bbC.\label{egjo11Dec23}
\end{align}
The Legendre function of the first kind can be defined by the so-called
Schl\"afli integral
\begin{align}
 P_\nu(z)&=
{1\over 2\pi i}\oint_C {dt\over t-z}\left[{t^2-1\over 2 (t-z)}\right]^{\nu}.\label{flxk12Jul23}
\end{align}
For general $\nu$, the integrand has logarithmic branch cuts which are
taken to be the intervals $[-\infty,-1]$ and $[1,z]$. The contour $C$
goes counterclockwise around the branch cut $[1,z]$.  The phase is
chosen so that $\arg({t^2-1\over t-z})\to 0$ for sufficiently large
$t>0$.  Thus defined $P_\nu(z)$ is single-valued on a $z$-plane with a
cut along $[-\infty,-1]$ and satisfies $P_\nu(1)=1$.

There are multiple ways to define the Legendre function of the second
kind, $Q_\nu(z)$, which is an independent solution of the Legendre
differential equation.  Here we define it as
\begin{align}
 Q_\nu(z)&={1\over 4i\sin{\pi\nu }}\oint_C {dt\over z-t}\left[{t^2-1\over 2 (z-t)}\right]^{\nu},\qquad \nu\in\bbC,
\label{ghnh31May23}
\end{align}
with the following contour:
\begin{align}
 \raisebox{-5ex}{
  \begin{tikzpicture}[scale=1.5]
 \node () at (-0.5,0.5) {$C$};
 \draw[fill=black] (0.3,0.9) circle (0.03) node [right] {$z$};
 \draw[fill=black] ( 1,0) circle (0.03) node [below] {$1$};
 \draw[fill=black] (-1,0) circle (0.03) node [below] {$-1$};
\draw [thick] plot [smooth cycle, tension=0.6] coordinates 
   { (0,0) (1,0.5) (1.5,0) (1,-0.5) (0,0) (-1,0.5) (-1.5,0) (-1,-0.5)};
\draw [-latex] (1.05,0.5) -- +(0.01,0);
\draw [-latex] (-1.05,0.5) -- +(-0.01,0);
\node () at (0,0) [below] {$0$};
\draw (1.45,0) -- (1.55,0) node [right,xshift=-2] {P};
\end{tikzpicture}}
\end{align}
The phase of the integrand is chosen so that $\arg(t^2-1)=0$ at
point P and $\arg(z-t)=\arg z$ at $t=0$.  Thus defined
$Q_\nu(z)$ has a branch cut along $[-\infty,1]$ and defined for all
$\nu\in\bbC$ except for $\nu=-1,-2,\cdots$.  In Mathematica, this
$Q_\nu(z)$ is called the Legendre function of the second kind of type 3.
For $\Re(1+\nu)>1$, this $Q_\nu(z)$ can  be rewritten as
\begin{align}
 Q_\nu(z)={1\over 2}\int_{-1}^1 {(1-t^2)^\nu\over 2^\nu (z-t)^{\nu+1}}dt.
 \label{ehpf11Jul23}
\end{align}

The explicit expressions for some small values of $\nu$ are
\begin{align}
\begin{split}
 P_0(z)&=1,\qquad
 P_1(z)=z,\qquad
 P_2(z)={1\over 2}(3z^2-1),\\[1ex]
 Q_0(z)&={1\over 2}\log{z+1\over z-1},\qquad
 Q_1(z)={1\over 2}P_1(z)\log{z+1\over z-1}-1,\\
 Q_2(z)&={1\over 2}P_2(z)\log{z+1\over z-1}-{3z\over 2}.
\end{split}
\end{align}

Because the Legendre equation \eqref{egjo11Dec23} is unchanged under
$\nu\to -\nu -1$, the Legendre functions $(P_{\nu-1/2},Q_{\nu-1/2})$ are
linear combinations of $(P_{-\nu-1/2},Q_{-\nu-1/2})$.  Indeed,
$P_\nu(z)$ satisfies
\begin{align}
 P_{\nu-1/2}(z)=P_{-\nu-1/2}(z),\qquad \nu\in\bbC,
 \label{epvq12Jul23}
\end{align}
while $Q_\nu(z)$ satisfies
\begin{align}
 Q_{m-1/2}(z)=Q_{-m-1/2}(z),\qquad m\in\bbZ.\label{jwgm24Jul23}
\end{align}

The behavior near $z=1$ is
\begin{subequations} 
 \begin{align}
 P_\nu(z)&=1+{\nu(1+\nu)\over 2}(z-1)+\cdots,\\
 Q_\nu(z)&=-{1\over 2}\log{z-1\over 2}
 -(\psi(\nu+1)+\gamma)+\cdots,
\label{ncxa11Dec23}
 \end{align}
\end{subequations}
where $\psi(z)$ is the digamma function and $\gamma$ is the
Euler-Mascheroni constant. The behavior near $z=\infty$ is, for generic
$\nu\notin\bbZ/2$,
\begin{align}
 P_\nu(z)
&={2^{-\nu-1}\Gamma(-\nu-1/2)\over \sqrt{\pi}\,\Gamma(-\nu)}(z-1)^{-\nu-1}\left(1-{\nu+1\over z-1}+\cdots\right)
\notag\\
&\quad+{2^{\nu}\Gamma(\nu+1/2)\over \sqrt{\pi}\,\Gamma(\nu+1)}(z-1)^{\nu}\left(1+{\nu\over z-1}+\cdots\right).
\end{align}
If $\nu\in\bbR$ and $\nu\notin \bbZ+\half$,
 \begin{align}
 P_\nu(z)=
\begin{cases}
  \displaystyle {\Gamma(2\nu+1)\over 2^\nu \Gamma(\nu+1)^2}z^{\nu}+\cdots & (\nu>-{1\over 2})\\[3ex]
  \displaystyle    {2^{\nu+1}\Gamma(-2\nu-1)\over \Gamma(-\nu)^2}z^{-\nu-1}+\cdots &  (\nu<-{1\over 2})\\
\end{cases} 
\label{lmnz11Dec23}
\end{align}
So, as $z\to \infty$, $P_\nu(u)$ diverges for $|\nu-1/2|>0$ and vanishes
for $|\nu-1/2|< 0$.
The expansion for
$Q_\nu(z)$ for generic $\nu$ is
\begin{align}
 Q_\nu(z)
 ={2^{-\nu - 1} \sqrt{\pi}\, \Gamma(\nu + 1)\over \Gamma(\nu+3/2)}z^{-\nu-1}\left[
 1+{(1+\nu)(2+\nu)\over 2(3+2\nu)z^2}
 +\cdots
 \right].\label{jwhh24Jul23}
\end{align}
For $\nu> -1$, this vanishes as $z\to \infty$
 while, for generic $\nu<-1$, it diverges as $z\to \infty$.

By the coordinate transformation \eqref{mgha6Nov23} with $a=2$,
$P_\nu(z)$ can be written as
\begin{align}
 P_\nu(z)={1\over 2\pi i}&\oint_C {w^\nu dw\over \sqrt{w^2-2zw+1}}
 ={1\over \pi}\int_{w_-}^{w_+} {w^\nu dw\over \sqrt{-w^2+2zw-1}}
 \label{ehpq11Dec23}
\end{align}
where in the first expression the contour goes counterclockwise around
the cut $[w_-,w_+]$, $w_\pm=u\pm \sqrt{u^2-1}$, and the phase is taken
so that $\arg(w^2-2zw+1)=0$ for positive $w>w_+$.  Likewise, $Q_\nu(z)$
can be written as
\begin{align}
 Q_\nu(z) = \int_0^{w_-} {w^\nu dw\over \sqrt{w^2-2zw+1}}.\label{ehwe11Dec23}
\end{align}
We could define this as a contour integral along a Pochhammer
contour going around $w=0,w_-$.

\subsection{Resonant Legendre functions}

The resonant Legendre equation \cite{Backhouse:1986} is the Legendre
equation \eqref{egjo11Dec23} with a inhomogeneous term proportional to
the Legendre function.  Here we consider the following equation in particular:
\begin{align}
 (1-z^2)\cP_\nu''(z)-2z \cP_\nu'(z)+\nu (\nu+1)\cP_\nu(z)=-(2\nu+1)P_\nu(z).\label{msym22Dec23}
\end{align}
By differentiating the Legendre differential equation
\eqref{egjo11Dec23} with respect to $\nu$, we immediately see that the
solution $\cP_\nu(z)$ is nothing but the $\nu$-derivative of $P_\nu(z)$:
\begin{align}
 \cP_\nu(z)=\p_\nu P_\nu(z)
 ={1\over 2\pi i}\oint_C{w^\nu \log w\,dw\over
 \sqrt{1-2zw+w^2}}
 ={1\over \pi}\int_{w_-}^{w_+}{w^\nu \log w\,dw\over
 \sqrt{-1+2zw-w^2}},
 \label{ehtf11Dec23}
\end{align}
where we used \eqref{ehpq11Dec23}.  Similarly, we can define
$\cQ_\nu(z)$ as the
solution to the differential equation \eqref{msym22Dec23} with
$P_\nu(z)$ on the right hand side replaced by $Q_\nu(z)$. From
\eqref{ehwe11Dec23}, we see that it can be written as
\begin{align}
 \cQ_\nu(z) 
 =\p_\nu Q_\nu(z) 
 = \int_0^{w_-} {w^\nu \log w\,dw\over \sqrt{w^2-2zw+1}}.
\end{align}

The explicit expressions for some small values of $\nu$ are
\begin{align}
\begin{split}
   \cP_0(z)&=\log{z+1\over 2},\qquad
 \cP_1(z)=z\log{z+1\over 2}+z-1,\\
 \cP_2(z)&=P_2(z)\log{z+1\over 2}+ {7\over 4}z^2-{3\over 2}z-{1\over 4},\\[2ex]
  \cQ_0(z)&=-{1\over 2}\log{z+1\over 2}\log{z-1\over 2}-{\rm Li}_2{1-z\over 2}-{\pi^2\over 6},\\
 \cQ_1(z)&=z\cQ_0(z)
 +{z+1\over 2}\log{z+1\over 2} -{z-1\over 2}\log{z-1\over 2}+1,
\\
 \cQ_2(z)&=P_2(z)\cQ_0(z)
 +\left(-\frac{7 z^2}{8}+\frac{3 z}{4}+\frac{1}{8}\right) \log  \left(\frac{z-1}{2}\right)
\\
 &\qquad\qquad\qquad
 +\left(\frac{7 z^2}{8}+\frac{3 z}{4}-\frac{1}{8}\right) \log  \left(\frac{z+1}{2}\right)+\frac{5 z}{4}.
\end{split}
\end{align}
For $\nu=-1/2$, the resonant Legendre equation is the same as the
Legendre equation and therefore the former must be expressible in terms
of the latter.  Indeed,
\begin{align}
\begin{split}
  \cP_{-1/2}(z)&=0,\\
 \cQ_{-1/2}(z)&=-{\pi^2\over 2}P_{-1/2}(z)
 =-\pi {\bf K}(\tfrac{1-z}{2}),
\end{split}\label{mryq11Dec23}
\end{align}
where ${\bf K}(m)$ is the complete elliptic integral of the first kind.

The values at $z=1$ are
\begin{align}
 \cP_\nu(z=1)=0,\qquad
 \cQ_\nu(z=1)=-\psi^{(1)}(1+\nu),
\end{align}
where $\psi^{(m)}(x)$ is the polygamma function.

The behavior as $z\to \infty$ is, for $\nu\in\bbR$ and $\nu\notin \bbZ+\half$,
 \begin{align}
 \cP_\nu(z)=
\begin{cases}
  \displaystyle {\Gamma(2\nu+1)\over 2^\nu \Gamma(\nu+1)^2}z^{\nu}\log z+\cdots & (\nu>-{1\over 2})\\[3ex]
  \displaystyle  -{2^{\nu+1}\Gamma(-2\nu-1)\over \Gamma(-\nu)^2}z^{-\nu-1}\log z+\cdots &  (\nu<-{1\over 2})\\
\end{cases}\label{mbxu11Dec23} 
\end{align}
The behavior of $\cQ_\nu(z)$ as $z\to \infty$ is, for generic $\nu$,
\begin{align}
 \cQ_\nu(z)
 = -{2^{-\nu - 1} \sqrt{\pi}\, \Gamma(\nu + 1)\over \Gamma(\nu+3/2)}z^{-\nu-1}\log z+\cdots.\label{mbyh11Dec23}
\end{align}
For $\nu> -1$, this vanishes as $z\to \infty$
while, for generic $\nu<-1$, it diverges as $z\to \infty$.

\subsection{Relation to harmonic functions}

As discussed in the main text,
\begin{align}
 \sqrt{u-\cos\sigma}\,e^{\pm i(\nu+\half)\sigma} P_\nu(u),
 \,\qquad
 \sqrt{u-\cos\sigma}\,e^{\pm i(\nu+\half)\sigma} Q_\nu(u)
\end{align}
are harmonic functions in 3D\@.  If $\nu>-\half$ and $\nu \notin
\bbZ+\half$, the $u\to\infty$ behavior is, from \eqref{lmnz11Dec23},
\begin{subequations} 
\label{mtgd22Dec23}
 \begin{align}
 \sqrt{u-\cos\sigma}\,e^{\pm i(\nu+\half)\sigma} P_\nu(u)
 &\sim
  (e^{\pm i\sigma}u)^{\nu+\half}
 \propto \begin{cases}\zb^{-\nu -\half}\\ z^{-\nu -\half}\end{cases},
  \label{lmzw11Dec23}
 \\[1ex]
 \sqrt{u-\cos\sigma}\,e^{\pm i(\nu+\half)\sigma} Q_\nu(u)
 &\sim
 \Bigl({e^{\pm i\sigma}\over u}\Bigr)^{\nu+\half}
 \propto \begin{cases}z^{\nu+\half}\\ \zb^{\nu+\half}\end{cases},
  \label{lnec11Dec23}
 \end{align}
\end{subequations}
where we also wrote down the corresponding 2D functions using the
identification $z\propto e^{i\sigma}/u$ (Eq.~\eqref{gxay19Apr17}).
\eqref{lmzw11Dec23} corresponds to (anti)holomorphic functions that
diverge near the 2D infinity, while \eqref{lnec11Dec23} corresponds to
(anti)holomorphic functions that vanish near the 2D infinity.

As discussed in \eqref{jscc11Dec23}, the combinations
\begin{align}
\begin{gathered}
  \sqrt{u-\cos\sigma}\, e^{\pm i(\nu+\half)\sigma}(\cP_\nu(u) \pm i\sigma P_\nu(u)),\\
 \sqrt{u-\cos\sigma}\, e^{\pm i(\nu+\half)\sigma}(\cQ_\nu(u) \pm i\sigma Q_\nu(u))
\end{gathered}
 \label{jvfm11Dec23}
\end{align}
are also harmonic functions in 3D\@. Another way to see the harmonicity
is as follows. Using the integral expressions \eqref{ehpq11Dec23},
\eqref{ehtf11Dec23}, this can be rewritten as
\begin{align}
 \sqrt{u-\cos\sigma}\,& e^{\pm i(\nu+\half)\sigma}(\cP_\nu(u) \pm i\sigma P_\nu(u))
\notag\\
 &=\sqrt{u-\cos\sigma}\, e^{\pm i\sigma/2} 
 {1\over 2\pi i} \oint {(w e^{\pm i\sigma})^\nu \log (we^{\pm i\sigma})\,dw\over \sqrt{w^2-2uw+1}}\notag\\
 &=\sqrt{u-\cos\sigma}\, e^{\pm i\sigma/2} 
 {1\over 2\pi i} \oint {w'^\nu \log w'\,dw'\over \sqrt{w'^2-2uw'e^{\pm i\sigma}+e^{\pm 2i\sigma}}}
\end{align}
where we set $we^{\pm i\sigma}=w'$. This is harmonic,
because the Laplacian of
\begin{align}
 \sqrt{u-\cos\sigma}\,e^{\pm i\sigma/2}
 {1\over \sqrt{w'^2-2uw'e^{\pm i\sigma}+e^{\pm 2i\sigma}}}
\end{align}
identically vanishes for any $w'$.  The same is true for the
combination of $Q,\cQ$ in \eqref{jvfm11Dec23}.

The behavior of \eqref{jvfm11Dec23} as $u\to \infty$ for $\nu >\half$
and $\nu \notin\bbZ+\half$ is, from \eqref{mbxu11Dec23} and
\eqref{mbyh11Dec23}, 
\begin{subequations}
 \begin{align}
 \sqrt{u-\cos\sigma}\,e^{\pm i(\nu+\half)\sigma} (\cP_\nu(u)\pm i\sigma P_\nu(u))
 &\sim
  (e^{\pm i\sigma}u)^{\nu+\half} \log (e^{\pm i\sigma}u)
 \sim
  \begin{cases}
  \zb^{-\nu -\half}\log\zb  \\ 
  z^  {-\nu -\half}\log z
  \end{cases},
  \\[1ex]
 \sqrt{u-\cos\sigma}\,e^{\pm i(\nu+\half)\sigma} (\cQ_\nu(u)\pm i\sigma Q_\nu(u))
 &\sim
  \Bigl({e^{\pm i\sigma}\over u}\Bigr)^{\nu+\half}\log\Bigl({e^{\pm i\sigma}\over u}\Bigr)
 \sim
  \begin{cases}
  z^{\nu +\half}\log z  \\ 
  \zb^{\nu +\half} \log\zb
  \end{cases}.
\label{mpck11Dec23}
 \end{align}
\end{subequations}

\section{The ordinary supertube}
\label{app:supertube}

In the main text, we studied 3D solutions by extending 2D solutions that
involve multiple stacks of branes with non-Abelian monodromies (the
constant-$\tau$ solutions are regarded as multiple stacks with
non-Abelian monodromies collapsing to a point).  However, a simpler
possibility is to start with the 2D solution corresponding to a single
brane,
\begin{align}
 f={1\over 2\pi i}\log{z\over \mu},\qquad g=1,\label{hniw11Dec23}
\end{align}
where $\mu$ is a constant.
As we go around $z=0$, this  solution has the monodromy
\begin{align}
 \mqty(f\label{hqdw11Dec23}\\g) \to \mqty(1&1\\ 0&1)\mqty(f\\ g).
\end{align}
The 2D solution \eqref{hniw11Dec23} physically makes sense only near the
origin; for $|z|>\mu$, $\Im\tau=\Im(f/g)={1\over 2\pi}\log{\mu\over
|z|}$ gets negative and the solution is not acceptable.  The point of
F-theory is to resolve this problem at long distance by including other
branes with non-commuting monodromies. However, if we extend
\eqref{hniw11Dec23} to 3D, it is possible that this is resolved by a
different long distance effect, namely, the infinitely long straight
brane in 2D is actually a closed circle in 3D\@.

Indeed, the 3D solution that reduces to \eqref{hniw11Dec23} in the 2D
limit is just the ordinary supertube and the  harmonic functions are
given by \cite{deBoer:2012ma}
\begin{align}
 F=\gamma+i\varphi,\qquad G=1,\label{iphs11Dec23}
\end{align}
where
\begin{align}
 \varphi(u,\sigma)&={1\over 2\sqrt{2}}\sqrt{u-\cos\sigma}\,P_{-1/2}(u),\label{mrra11Dec23}\\
 \gamma(u,\sigma)&={1\over R}
\left[
 {a(u)\over\sqrt{u-1}}\,{\bf F}(\tfrac{\sigma}{2}|{-\tfrac{2}{u-1}})
 -2\sqrt{u-1}\,a'(u)\, {\bf E}(\tfrac{\sigma}{2}|{-\tfrac{2}{u-1}})
 \right],\label{dpfm25Dec23}
 \\
 a(u)&=
-{R\over 8\sqrt{2}}
 {u^2-1\over u^{3/2}}\,\,_2F_1\left(\tfrac{3}{4},\tfrac{5}{4};2; 1-u^{-2}\right).
\end{align}
Here, ${\bf F}(\phi|m)$ and ${\bf E}(\phi|m)$ are the elliptic integrals
of the first and second kinds, respectively.  This solution has
$\Im\tau=\varphi(u,\sigma)>0$ everywhere and has the monodromy
\eqref{hqdw11Dec23} because
\begin{align}
 \gamma(u,\sigma) \to \gamma(u,\sigma) + 1\qquad \text{as}\qquad \sigma \to \sigma+2\pi,\label{jggz11Dec23}
\end{align}
as can be shown from the properties of the elliptic
integrals. Furthermore, the $u\to \infty$ behavior is
\begin{align}
 \gamma&={\sigma\over 2\pi }+\cO(u^{-1}),\qquad
 \varphi={\log(8u)\over 2\pi}+\cO(u^{-1}),\notag\\
 F&
 ={1\over 2\pi i}\log{e^{i\sigma}\over 8u} +\cO(u^{-1})\label{mrfg11Dec23}
\end{align}
which, with the relation \eqref{iepw1Mar23}, reduces to
\eqref{hniw11Dec23} with an appropriate choice for $\mu$.
Also, the value at $u= 1$ is 
\begin{align}
 \gamma(u=1,\sigma)=
 \sin^2\left( {\pi\over 2}\left\{{\sigma\over 2\pi}\right\}\right)
 +\left\lfloor {\sigma\over 2\pi} \right\rfloor,
\label{nhjf11Dec23}
\end{align}
where $\{x\}\equiv x-\lfloor x\rfloor$ is the fractional part of $x$ and
$\lfloor x\rfloor$ is the floor function.  This is a smooth function
with its second derivative being discontinuous at $\sigma\in 2\pi\bbZ$.

A natural question is whether we can obtain the 3D harmonic function
\eqref{iphs11Dec23} upon applying the extension formula
\eqref{extension_formula}.  It turns out that that is not possible,
because the expansion of~$F$ contains only $Q_\nu$ and $\cQ_\nu$, while
the extension formula \eqref{extension_formula} only contains $P_\nu$.
Below we study this in more detail.

\subsection{Partition of unity}
\label{sss:partition_of_unity}

One possible starting point is to ask the following question. Obviously
\begin{align}
 1=\sqrt{u-\cos\sigma}\cdot {1\over \sqrt{u-\cos\sigma}}.
\end{align}
From the discussion around \eqref{iqhz11Dec23}--\eqref{gmyq11Dec23},
we must be able to expand $(u-\cos\sigma)^{-1/2}$ in terms of
$P_\nu(u)$ and $Q_\nu(u)$.  What is the expansion?

By expanding $(u-\cos\sigma)^{-1/2}$ in $u^{-1}$ and
collecting terms with the same power of $e^{i\sigma}$, it is
straightforward to show
\begin{align}
 {1\over \sqrt{u-\cos\sigma}}
 &=\sum_{m=-\infty}^{\infty}
 e^{im\sigma} {\Gamma(|m|+\half)\over  \Gamma(|m|+1)}
 { {}_2F_1({1\over 4}+{|m|\over 2},{3\over 4}+{|m|\over 2};1+|m|;u^{-2})
 \over \sqrt{\pi u\,}(2u)^{|m|}}.
\label{mqmi10Jul23}
\end{align}
From the discussion around \eqref{iqhz11Dec23}--\eqref{gmyq11Dec23} and
from the relations \eqref{epvq12Jul23} and \eqref{jwgm24Jul23}, the
coefficient of $e^{im\sigma}$ must be expressible as a linear
combination of $P_{|m|-1/2}(u)$ and $Q_{|m|-1/2}(u)$.  We can determine
the linear combination by looking at the behavior as $u\to \infty$.
Because $_2F_1(a,b;c;z)\to 1$ as $z\to 0$, the coefficient goes like
$u^{-|m|-1/2}$.   As $u\to \infty$, 
the behavior of $P_{|m|-1/2}(u)$, $Q_{|m|-1/2}(u)$ is
\begin{align}
 P_{|m|-1/2}(u)&=
\begin{cases}
 \cO(u^{-1/2}, u^{-1/2}\log u) & (|m|=0)\\
 \cO(u^{|m|-1/2}) & (|m|\ge 1)
\end{cases},
 \\
 Q_{|m|-1/2}(u)&={\sqrt{\pi}\,\Gamma(|m|+1/2)\over \Gamma(|m|+1)}{1\over (2u)^{|m|+1/2}}+\cdots.
\end{align}
So, the coefficients in \eqref{mqmi10Jul23} must contain only
$Q_{|m|-1/2}(u)$ and no $P_{|m|-1/2}(u)$.  After fixing the coefficients by
comparison, we find that
\begin{align}
 {1\over \sqrt{u-\cos\sigma}}
 &=
  {\sqrt{2}\over \pi}\sum_{m=-\infty}^\infty e^{im\sigma}\,Q_{|m|-1/2}(u).
 \label{nhqt11Jul23}
\end{align}
In other words,
\begin{align}
 1
 &=
  \sqrt{u-\cos\sigma}\cdot
 {\sqrt{2}\over \pi}
 \biggl[Q_{-1/2}(u)
 +\sum_{m=1}^\infty \Bigl(e^{im\sigma}Q_{m-1/2}(u)+e^{-im\sigma}Q_{m-1/2}(u)\Bigr)
 \biggr].
\label{nexo10Jul23}
\end{align}
This is how we expand unity in terms of the basis of harmonic functions,
$P,Q$.

\subsection{The supertube harmonic function expanded in $Q,\cQ$}

According to \eqref{jfbq11Dec23},
\begin{align}
H= {\sqrt{u-\cos\sigma}\over 2\pi i}  \,
 e^{\pm im\sigma}( \cQ_{m-1/2} \pm i\sigma Q_{m-1/2})
\end{align}
is a harmonic function that has the monodromy
\begin{align}
H\to H \pm \sqrt{u-\cos\sigma}\, e^{\pm im\sigma} Q_{m-1/2}.
\end{align}
Combining this fact with the relation \eqref{nexo10Jul23}, we conclude that
$\gamma$, which is harmonic and has the monodromy \eqref{jggz11Dec23},
must be given by
\begin{align}
 \gamma(u,\sigma)
 &={\sqrt{u-\cos\sigma}\over 2\pi i}{\sqrt{2}\over \pi}
 \biggl[a(\cQ_{-1/2}+i\sigma Q_{-1/2})
 -b(\cQ_{-1/2}-i\sigma Q_{-1/2})
 \notag\\
 &\quad
 +\sum_{m=1}^\infty \Bigl(e^{im\sigma}(\cQ_{m-1/2}+i\sigma Q_{m-1/2})
 -e^{-im\sigma}(\cQ_{m-1/2}-i\sigma Q_{m-1/2})\Bigr)\biggr]\label{mpdn11Dec23}
\end{align}
with $a+b=1$.  The values of $a,b$ cannot be determined only by
monodromy.
From \eqref{mpck11Dec23}, the leading $u\to \infty$ behavior of
\eqref{mpdn11Dec23} is
\begin{align}
 \gamma
\sim (a-b)(i\sigma - \log u).
\end{align}
By comparing this with the desired behavior of $\gamma$,
\eqref{mrfg11Dec23}, we find that we must take $a=b=1/2$.  It is not
difficult to check that, with this choice, \eqref{mpdn11Dec23} is indeed
equal to \eqref{dpfm25Dec23}.

On the other hand, using \eqref{mryq11Dec23}, the function $\varphi$ in
\eqref{mrra11Dec23} can be written as
\begin{align}
 \varphi &= -{1\over \sqrt{2}\pi^2}\sqrt{u-\cos\sigma }\, \cQ_{-1/2}
 \notag\\
 &= -{1\over 2\sqrt{2}\pi^2}\sqrt{u-\cos\sigma }\, 
 [(\cQ_{-1/2}+i\sigma Q_{-1/2})+(\cQ_{-1/2}-i\sigma Q_{-1/2})].
\end{align}
Combining this with \eqref{mpdn11Dec23}, we finally find the expression
for the 3D harmonic function $F=\gamma+i\varphi$ in terms of $Q,\cQ$:
\begin{align}
 F
 &={\sqrt{u-\cos\sigma}\over 2\pi i}{\sqrt{2}\over \pi}
 \biggl[(\cQ_{-1/2}+i\sigma Q_{-1/2})
 \notag\\
 &\qquad\qquad
 +\sum_{m=1}^\infty \Bigl(e^{im\sigma}(\cQ_{m-1/2}+i\sigma Q_{m-1/2})
 -e^{-im\sigma}(\cQ_{m-1/2}-i\sigma Q_{m-1/2})\Bigr)\biggr].
\label{hqvk13Dec23}
\end{align}

Some comments are in order.  Using \eqref{mpck11Dec23}, we see that the
first line of \eqref{hqvk13Dec23} tends to the holomorphic term $\log z$
in the 2D limit, which reproduces the desired 2D
expression~\eqref{hniw11Dec23}.  Therefore, this line is required by
matching.  On the other hand, the two terms in the second line tend to
$z^m \log z$ and $\zb^m \log\zb$. These do not correspond to anything in
the 2D harmonic function $f$; because they vanish in the 2D limit, $u\to
\infty$ ($|z|\to 0$), their existence cannot be deduced from
matching. The second term is even anti-holomorphic and guessing its
existence from the holomorphic $f(z)$ is far from obvious.  However,
when extending $f$ to 3D, these terms are necessary for the monodromy
$\gamma\to \gamma+1$ in the entire $\bbR^3$, as we saw above.  Another
interesting point is that, because $Q_{m-1/2}(u)$ goes as ${1\over
2}\log{2\over u-1}$ as $u\to 1$ (see \eqref{ncxa11Dec23}), it seems that
$F$ is singular at $u=1$.  However, actually, the $u\to 1$ limit is
regular except at $\sigma\in 2\pi \bbZ$ as can be seen
from~\eqref{nhjf11Dec23}. Indeed, the sum of the leading contributions
is
\begin{align}
 \sum_{m\in\bbZ} e^{im\sigma} Q_{m-1/2}(u)
 &\sim
 {1\over 2}\log{2\over u-1}
 \sum_{m\in\bbZ} e^{im\sigma}
 =
 {1\over 2}\log{2\over u-1}\cdot
 2\pi \sum_{k\in\bbZ} \delta (\sigma - 2\pi k).
\end{align}
For $\sigma\notin 2\pi\bbZ$, this vanishes and leads to a regular $u\to 1$ limit.
For  $\sigma\in 2\pi\bbZ$, 
we have a $\delta$-function
singularity,
but that is consistent with the
fact that \eqref{nhjf11Dec23} is non-analytic at
$\sigma\in 2\pi\bbZ$.

\section{More examples}
\label{app:more_ex}

Here we present more examples of 2D harmonic functions to which we can
apply the extension formula.  In these examples, it turns out that the
monodromy structure in 3D is different from that in 2D, which is
puzzling.  When we look for genuine microstates of black holes that
involve codimension-2 sources, these examples should serve as useful
data points.

\subsection{Constant-$\tau$ solutions with two stacks}

In section \ref{sss:single_stack}, we studied 3D solutions based on the
constant-$\tau$ solution in 2D with a single stack of $N$ branes.  Here,
we consider 3D solutions based on the constant-$\tau$ solution in 2D
with two stacks of $N/2$ branes ($N$ branes in total).

As the 2D solution, let us consider the situation where 
there are the two stacks (each of $N/2$ branes) at $z=0,1$.
The harmonic functions are
\begin{align}
 g= {1\over (z(z-1))^{N/24}}, \qquad f=\tau_0 g,\label{mszh12Dec23}
\end{align}
where $\tau_0$ is constant.  The $z^{-1}$ expansion of $g$ is
\begin{align}
 g=\sum_{k=0}^\infty {\Gamma({N\over 24}+k)\over \Gamma({N\over 24})}
 {1\over k!}{1\over z^{k+{N\over 12}}}.
\label{lmad15Sep23}
\end{align}
By a straightforward application of the extension formula
\eqref{extension_formula}, we find that the 3D harmonic function is
\begin{align}
 G
 &= 
   {\Gamma({N\over 12}+{1\over 2})^2\over \Gamma({N\over 6})}
\sqrt{u-\cos\sigma\over Z}
 {1\over 2\pi i}\oint {dt\over t-u}
 \left({t^2-1\over (t-u)Z}\right)^{{N\over 12}-{1\over 2}}
 {}_2F_1\biggl({N\over 24},{N\over 12}+{1\over 2}; {N\over 12}; {t^2-1\over 4(t-u)Z}\biggr).\label{ffow15Jul23}
\end{align}
If we go to the $w$ coordinate using \eqref{mgha6Nov23} with $a=4$, this
can be written as
\begin{align}
 G
 &= 
   {\Gamma({N\over 12}+{1\over 2})^2\over \Gamma({N\over 6})}
\sqrt{u-\cos\sigma\over Z}
 {1\over 2\pi i}\oint_C {(4w)^{{N\over 12}-{1\over 2}}\,dw\over\sqrt{w^2-{u\over Z} w+{1\over 4Z^2}}}
 \,{}_2F_1\biggl({N\over 24},{N\over 12}+{1\over 2}; {N\over 12}; w\biggr).
\label{ihsk19Sep23}
\end{align}
The hypergeometric function $_2F_1(\alpha,\beta;\gamma;z)$ has a branch
cut along $[1,\infty]$.  So, the cut structure and the contour $C$ on
the $w$-plane are as below:
\begin{align}
  \tikzset{cross/.style={thick,cross out,draw=black,
  minimum size=5pt,inner sep=0pt, outer sep=0pt}}
\raisebox{-7ex}{
\begin{tikzpicture}[scale=1.5]
  \node (minfty) at (-1,0) {};
  \node (zero)  at (0,0) [circle,fill=black,inner sep=0,minimum size=3pt,label=above:$0$] {};
  \node (wm)    at (1,0) [circle,fill=black,inner sep=0,minimum size=3pt,label=above:$w_-$] {};
  \node (wp)    at (2,0) [circle,fill=black,inner sep=0,minimum size=3pt,label=above:$w_+$] {};
  \node (one)   at (3,0) [circle,fill=black,inner sep=0,minimum size=3pt,label=above:$1$] {};
  \node (infty) at (4,0) {};
 \draw[thick] (minfty) -- (zero);
 \draw[thick] (wm) -- (wp);
 \draw[thick] (one) -- (infty);
  \draw (1.5,0) ellipse (1 and 0.5);
 \draw[-latex] (1.5,-0.5) -- +(0.01,0) node [below] {$C$};
 \end{tikzpicture}}
\end{align}

The position of branes corresponds to the values of $(u,\sigma)$ at
which some cuts collide.  We find that this can happen for $w_+=1$ and
$w_-=0$, which gives the following brane positions:
\begin{align}
 (u,\sigma)=(u_*,l),~(\infty,\text{any}),
\end{align}
where $u_*$ was defined in \eqref{def_u*}.

In 2D, as we go around a stack of branes, the harmonic function gets
multiplied by the phase $e^{-\pi i N/12}$ as can be seen from
\eqref{mszh12Dec23}. In 3D, what is the change in the contour when we go
around the stack of branes at $(u,\sigma)=(u_*,l)$?  By carefully
studying how the contour transforms, one finds that
\begin{align}
 C\to C+C'
\end{align}
where $C'$ is a Pochhammer contour:
\begin{align}
 \tikzset{cross/.style={thick,cross out,draw=black,
  minimum size=5pt,inner sep=0pt, outer sep=0pt}}
 \raisebox{-2ex}{
\begin{tikzpicture}[scale=1.8]
 \draw [blue,domain=-18:310] plot ({2+0.1*cos(\x)}, {0.1*sin(\x)});
 \draw [blue,domain=23:368] plot ({2+0.2*cos(\x)}, {0.2*sin(\x)});
 \draw [blue](2.1,-0.025) -- (2.81,-0.025);
 \draw [blue](2.195,0.025) -- (2.9,0.025);
 \draw [blue](2.18,0.075) -- (2.82,0.075);
 \draw [blue](2.06,-0.075) -- (2.94,-0.075);
 \draw [blue,domain=-131:170] plot ({3+0.1*cos(\x)}, {0.1*sin(\x)});
 \draw [blue,domain=-172:157] plot ({3+0.2*cos(\x)}, {0.2*sin(\x)});
 \draw [blue,-latex] (2.4,0.075) -- +(0.01,0);
 \draw [blue,-latex] (2.4,0.025) -- +(-0.01,0);
 \draw [blue,-latex] (2.5,-0.025) -- +(-0.01,0);
 \draw [blue,-latex] (2.7,-0.075) -- +(0.01,0);
 \node () at (2.5,0.25) [blue] {$C'$};
  \node (minfty) at (-1,0) {};
  \node (zero)  at (0,0) [circle,fill=black,inner sep=0,minimum size=3pt,label=above:$0$] {};
  \node (wm)    at (1,0) [circle,fill=black,inner sep=0,minimum size=3pt,label=above:$w_-$] {};
  \node (wp)    at (2,0) [circle,fill=black,inner sep=0,minimum size=3pt,label={above,yshift=5}:$w_+$] {};
  \node (one)   at (3,0) [circle,fill=black,inner sep=0,minimum size=3pt,label={above,yshift=7}:$1$] {};
  \node (infty) at (4,0) {};
 \draw[thick] (minfty) -- (zero);
 \draw[thick] (wm) -- (wp);
 \draw[thick] (one) -- (infty);
 \end{tikzpicture}}
\end{align}
In general, the contour integral along $C+C'$ is not equal to $e^{-\pi i
N/12}$ times the contour integral along $C$.  This is easiest to see in
the $u\to 1$ limit, in which $w_+=w_-={1\over 2Z}$.  In this limit, the
contour integral along $C'$ diverges because it goes through the cut
that collapses.  This means that, if we approach the $x_3$-axis (where
$u=1$) avoiding the branes, $G$ is finite, but if go through the branes
and approach the $x_3$-axis, $G$ diverges.  Therefore, the monodromy
around branes of the 3D solution is different from that of the 2D
solution. We do not have a good physical understanding of this puzzling
result.

\subsection{Another non-constant $\tau$ solution with two stacks}

As another example, let us look at the curve
\begin{align}
 y^2=(x^2-\zh^2)(x-1).\label{ieih12Dec23}
\end{align}
Either by looking at the discriminant or by finding the values of $\zh$
at which branch points collide, we see that this setup contains a stack
of two branes at each of the points $\zh=-1,0,1$; namely, there are six branes in
total.  Unlike \eqref{fxax4Oct23}, there are no branes at $\zh=\infty$.
This curve can be obtained from \eqref{fxax4Oct23} by a fractional
linear transformation in $\zh$ to bring the brane at $\zh=\infty$ to
finite $\zh$.

We define the 2D harmonic functions as the period integrals of the torus
\eqref{ieih12Dec23},
\begin{align}
\begin{split}
 g(z)&= {1\over 2}\int_A {dx\over \sqrt{(x^2-z^2)(x-1)}}
 =\int_z^\infty {dx\over \sqrt{(x^2-z^2)(x-1)}}
 ={2{\bf K}({1+z\over 1-z})\over \sqrt{z-1}},\\
 f(z)&= {1\over 2}\int_B {dx\over \sqrt{(x^2-z^2)(x-1)}}
 =i\int_1^z {dx\over \sqrt{(z^2-x^2)(x-1)}}
 ={\sqrt{2}i{\bf K}({z-1\over 2z})\over \sqrt{z}}.
\end{split}\label{ibqp12Dec23}
\end{align}
The cuts and contours are taken as below:
 \begin{align}
 \tikzset{cross/.style={thick,cross out,draw=black,
  minimum size=5pt,inner sep=0pt, outer sep=0pt}}
 \raisebox{-7ex}
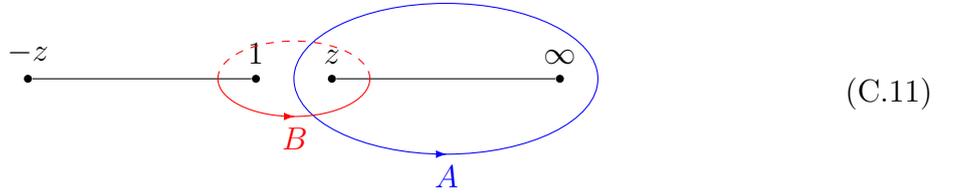
{
  \begin{tikzpicture}
\tikzset{
    partial ellipse/.style args={#1:#2:#3}{
        insert path={+ (#1:#3) arc (#1:#2:#3)}
    }
}
  \node (mz) at (-2,0)   [circle,fill=black,inner sep=0,minimum size=3pt,label=above:$-z$] {};
  \node (one) at (1,0)   [circle,fill=black,inner sep=0,minimum size=3pt,label=above:$1$] {};
  \node (z) at (2,0)     [circle,fill=black,inner sep=0,minimum size=3pt,label=above:$z$] {};
  \node (infty) at (5,0) [circle,fill=black,inner sep=0,minimum size=3pt,label=above:$\infty$] {};
 \draw[] (mz) -- (one);
 \draw[] (z) -- (infty);
 \draw[color=blue] (3.5,0) ellipse (2 and 1);
 \draw[-latex,color=blue] (3.5,-1) -- +(0.01,0) node [below] {$A$};
  \draw[dashed,color=red] (1.5,0) [partial ellipse=0:180:1 and 0.5];
  \draw[color=red] (1.5,0) [partial ellipse=180:360:1 and 0.5];
 \draw[-latex,color=red] (1.5,-0.5) -- +(0.01,0) node [below] {$B$};
  end angle = 150];
  \end{tikzpicture}
}
\label{laes1Feb23}
\end{align}
We define $\sqrt{(x^2-z^2)(x-1)}$ to be real and positive just below the
cut $[z,\infty]$.

As we did for \eqref{gnxc23Oct23}, we can find monodromy matrices as we
go around branes in the $\zh$ space, by following how the cycles
transform into each other. The result is:
\begin{align}
 M_{-1}=\begin{pmatrix} 1 & 2 \\ 0 & 1     \end{pmatrix},\quad
 M_0=\begin{pmatrix} -1 & 2 \\ -2 & 3     \end{pmatrix},\quad
 M_1=\begin{pmatrix} 1 & 0 \\ -2 & 1     \end{pmatrix},\quad
 M_{\infty}=\begin{pmatrix} -1 & 0 \\ 0 & -1     \end{pmatrix}.
 \label{jgso18Apr23}
\end{align}
These satisfy
\begin{align}
 M_1 M_0 M_{-1}=M_\infty.
\end{align}
The monodromy matrix $M_\infty=-{\bf 1}$ is consistent with what we said
below \eqref{iaht12Dec23}, because we are going around 6 branes; the
solution there can be thought of as the limit of the current
configuration in which all the branes collapse to a point.  We can also
consider going half around $z=\infty$ counterclockwise as $z\to
e^{i\pi}z$.  The associated monodromy matrix is
\begin{align}
 M_\infty^{1/2}=\begin{pmatrix} 0 & 1 \\ -1 & 0     \end{pmatrix}.
 \label{jgsi18Apr23}
\end{align}

We can derive the 3D harmonic functions as follows. By setting $x\to xz$
in \eqref{ibqp12Dec23} and expanding in $z^{-1}$, we find that
\begin{align}
 g(z)&=\int_1^\infty {dx\over \sqrt{(x^2-1)(zx-1)}}
 =\sum_{k=0}^\infty {\Gamma({1\over 2}+k)\over \Gamma({1\over 2})k!}{1\over z^{k+\half}}\int_1^\infty {x^{-k-{1\over 2}}dx\over\sqrt{x^2-1}}.\label{nhxw1Feb23}
\end{align}
Using the dictionary to go from a 2D harmonic function
\eqref{iuml1Mar23} to the 3D one \eqref{iupl1Mar23}, we find
\begin{align}
 G(u,\sigma)
 &=
 \sqrt{u-\cos\sigma\over Z}\sum_{k=0}^\infty
 \int_1^\infty dx{x^{-k-\half}\over\sqrt{x^2-1}}
 {P_k(u)\over (2Z)^k}\notag
\\
&=\sqrt{u-\cos\sigma\over Z}\int_1^\infty dx
 \sqrt{x\over (x^2-1)(x^2-{u\over Z}x+{1\over 4Z^2})},\label{iudu12Dec23}
\end{align}
where in the second equality we used the formula for the generating
function of the Legendre polynomials,
\begin{align}
 {1\over\sqrt{1-2tx+t^2}}=\sum_{k=0}^\infty P_k(x)t^k.
\label{llzm4Feb23}
\end{align}

What is peculiar about the result \eqref{iudu12Dec23} is the
appearance of the genus-two Riemann surface,
\begin{align}
 y^2=x(x^2-1)\left(x^2-{u\over Z}x+{1\over 4Z^2}\right).\label{ivir12Dec23}
\end{align}
We can regard $G$ in \eqref{iudu12Dec23} as a period of this Riemann
surface.  The roots of the right-hand side of \eqref{ivir12Dec23} are
\begin{align}
 x=0,\quad \pm 1,\quad x_\pm\equiv {1\over 2Z}\Bigl(u\pm \sqrt{u^2-1}\Bigr).
\end{align}
So, the branch cut structure of the genus-2 curve is as follows:
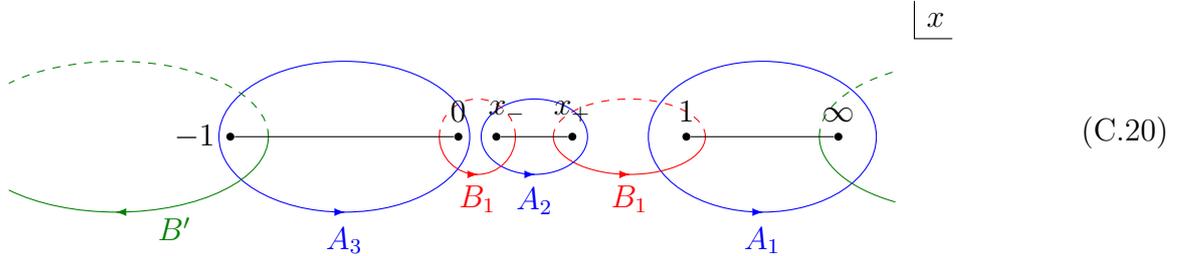
\begin{align}
 \raisebox{-9ex}{\begin{tikzpicture}
  \tikzset{cross/.style={thick,cross out,draw=black, minimum
  size=5pt,inner sep=0pt, outer sep=0pt}}
\tikzset{
    partial ellipse/.style args={#1:#2:#3}{
        insert path={+ (#1:#3) arc (#1:#2:#3)}
    }
}
 \draw (6.5,1.3) -- +(-0.5,0) -- +(-0.5,0.5) node [midway,xshift=8] {$x$};
 \draw[color=blue] (4,0) ellipse (1.5 and 1) node [below,yshift=-30] {$A_1$};
 \draw[-latex,color=blue] (4,-1) -- +(0.01,0);
 \draw[dashed,color=red] (2.25,0) [partial ellipse=0:180:1 and 0.5];
 \draw[color=red] (2.25,0) [partial ellipse=360:180:1 and 0.5] node [midway,below,yshift=0] {$B_1$};
 \draw[-latex,color=red] (2.25,-0.5) -- +(0.01,0);
 \draw[color=blue] (1,0) ellipse (0.7 and 0.5) node [below,yshift=-15] {$A_2$};
 \draw[-latex,color=blue] (1,-0.5) -- +(0.01,0);
 \draw[dashed,color=red] (0.25,0) [partial ellipse=0:180:0.5 and 0.5];
 \draw[color=red] (0.25,0) [partial ellipse=360:180:0.5 and 0.5] node [midway,below,yshift=0] {$B_1$};
 \draw[-latex,color=red] (0.25,-0.5) -- +(0.01,0);
 \draw[color=blue] (-1.5,0) ellipse (1.65 and 1) node [below,yshift=-30] {$A_3$};
 \draw[-latex,color=blue] (-1.5,-1) -- +(0.01,0);
 \draw[color=green!50!black] (-4.5,0) [partial ellipse=360:225:2 and 1] node [midway,below,yshift=0] {$B'$};
 \draw[color=green!50!black,dashed] (-4.5,0) [partial ellipse=0:135:2 and 1];
 \draw[-latex,color=green!50!black] (-4.5,-1) -- +(-0.01,0);
 \draw[color=green!50!black] (6.75,0) [partial ellipse=180:240:2 and 1];
 \draw[color=green!50!black,dashed] (6.75,0) [partial ellipse=180:120:2 and 1];
%
%
  \node (mZoveru) at (-3,0) [circle,fill=black,inner sep=0,minimum size=3pt,label=left:$-1$] {};
  \node (zero) at (0,0)     [circle,fill=black,inner sep=0,minimum size=3pt,label=above:$0$] {};
		  \node (xm) at (0.5,0)     [circle,fill=black,inner sep=0,minimum size=3pt,label=above:$\,\,\,\,x_-$] {};
  \node (xp) at (1.5,0)     [circle,fill=black,inner sep=0,minimum size=3pt,label=above:$x_+$] {};
  \node (Zoveru) at (3,0)   [circle,fill=black,inner sep=0,minimum size=3pt,label=above:$1$] {};
  \node (infty) at (5,0)    [circle,fill=black,inner sep=0,minimum size=3pt,label=above:$\infty$] {};
 \draw[] (mZoveru) -- (zero);
 \draw[] (xm) -- (xp);
 \draw[] (Zoveru) -- (infty);
 \end{tikzpicture}}
\end{align}
where we named various cycles.  These cycles are not independent but
satisfy
\begin{align}
 A_3=-A_1-A_2,\qquad B'=B_1+B_2.\label{ngzv1Jul23}
\end{align}
The contour used to define $G$ in \eqref{iudu12Dec23} is $A_1$, namely,
\begin{align}
 G=\oint_{A_1}\omega,\qquad \omega={1\over 2}\sqrt{u-\cos\sigma\over Z}\,{x\over y}dx.
\end{align}
As we
go from 2D to 3D, the cut on the left in \eqref{laes1Feb23} has split
into two cuts.

To better understand the situation, we would like to study the monodromy
as we go around branes.  We can determine the location of branes by
studying when the branch points collide. A short analysis tells us that
the branes are at
\begin{align}
 (u,\sigma)=(u_*,l),\,(u_*,l-\pi),\,(\infty,\text{any}),
\end{align}
where 
$u_*$ was defined in \eqref{def_u*} and
we assumed that $|Z|\ge 1/2$.  These correspond to the location of
the branes in the 2D solution, $\zh=1,-1,0$, respectively.  By carefully
following how the cycles transform into each other, we find the
following monodromy matrices:
\begin{align}
\begin{aligned}
 M_{-1}&=\mqty(1&2&2&0 \\ 0&1&0&0 \\ 0&0&1&0 \\ 0&-2&-2&1),&
 M_1&=\mqty(1&0&0&0 \\ 0&1&0&0 \\ -2&0&1&0 \\ 0&0&0&1),\\
 M_0&=\mqty(-1&0&2&0 \\ 0&1&0&0 \\ -2&0&3&0 \\ 0&2&0&1),&
 M_\infty&=\mqty(-1&-2&0&0 \\ 0&1&0&0 \\ 0&0&-1&0\\ 0&0&-2&1).
\end{aligned}\label{jevh12Dec23}
\end{align}
where we presented the matrices in the symplectic basis of
$(b_1,b_2,a_1,a_2)$, where
\begin{align}
 a_1=A_1,\qquad b_1=B_1,\qquad a_2=-A_3=A_1+A_2,\qquad b_2=B_2.
\end{align}
These matrices 
$Sp(4,\bbZ)$ are matrices satisfying
\begin{align}
  M^T J M=J,\qquad
 J=\mqty(0&{\bf 1}_2 \\ -{\bf 1}_2&0).
\end{align}
They also satisfy
\begin{align}
 M_1 M_0 M_{-1}=M_\infty.
\end{align}
We can also go halfway around all the branes counterclockwise, for which
the monodromy matrix is
\begin{align}
 M_\infty^{1/2}=\mqty(0&1&1&0\label{jezm12Dec23}\\ 0&-1&0&0\\ -1&-1&0&0\\ -1&-1&1&-1).
\end{align}
The 3D monodromy matrices \eqref{jevh12Dec23}, \eqref{jezm12Dec23}
correspond to the 2D monodromy matrices \eqref{jgso18Apr23},
\eqref{jgsi18Apr23}.  We can see that, if we take the submatrices of the
3D matrices by only keeping the $(b_1,a_1)$ part, we recover the
corresponding 2D matrices.

How should we define the other harmonic function $F$? We want the 3D
solution to have the same monodromy as the 2D solution. In 2D, for
example, as we go around the $\zh=1$ brane, cycle $A$ turns into $A-2B$,
as can be read off from \eqref{jgso18Apr23}.  In 3D, as we go around the
$(u_*,l)$ brane, the cycle $A_1=a_1$ goes to $a_1-2b_1=A_1-2B_1$.  So,
this suggests that we define~$F$ by
\begin{align}
 F\stackrel{?}{=}\oint_{B_1}\omega.
\end{align}
What happens to this $F$, if we go around the $(u_*,l-\pi)$ brane?  In
2D, as we go around the $\zh=-1$ brane, $B$ goes to $B+2A$ and thus $f$
goes to $f+2g$.  However, in 3D, the monodromy matrix
\eqref{jevh12Dec23} says that the cycle $B_1=b_1$ goes to
$b_1+2b_2+2a_1=B_1+2B_2+2A_1$.  This is not $F+2G$ but we have an extra
contribution, $2\oint_{B_2}\omega$.  Therefore, the monodromy structure
of the harmonic functions in 3D has changed from that in 2D\@!
In other words, although we started in 2D with two harmonic functions
$(f,g)$ that transform under $SL(2,\bbZ)$, we ended up in 3D with four
harmonic functions that transform under $Sp(4,\bbZ)$.  Relatedly, the
genus-one Riemann surface (torus) in 2D has become a genus-two Riemann
surface.  We do not have a good physical understanding of this puzzling
situation.


\end{document}